\documentclass[a4paper,10pt]{article}
\usepackage[utf8]{inputenc}
\usepackage{graphics} 
\usepackage{epsfig} 
\usepackage[retainorgcmds]{IEEEtrantools}
\usepackage{amsfonts,latexsym,amssymb} 
\usepackage{mathrsfs,amsmath}
\usepackage{framed}
\usepackage{upgreek}
\usepackage{bbm}
\usepackage{graphicx}
\usepackage{color}
\usepackage{tikz}
\usepackage{caption}
\usepackage{subcaption}
\usepackage{epstopdf}

\newcommand{\vecF}{\mathbf{F}}
\newcommand{\vecG}{\mathbf{G}}
\newcommand{\vecH}{\mathbf{H}}

\newcommand{\vecM}{\mathbf{M}}

\newcommand{\vecP}{\mathbf{P}}

\newcommand{\vect}{\mathbf{t}}

\newcommand{\vecX}{\mathbf{X}}

\newcommand{\vecf}{\boldsymbol{f}}
\newcommand{\vecg}{\boldsymbol{g}}
\newcommand{\vech}{\boldsymbol{h}}

\newcommand{\vecn}{\mathbf{n}}

\newcommand{\vecv}{\mathbf{v}}

\newcommand{\vep}{\varepsilon}

\newcommand{\vecphi}{\boldsymbol\upphi}

\newcommand{\vecphid}{\dot{\boldsymbol\upphi}}

\newtheorem{remark}{Remark}[section]

\begin{document}

\title{Energy-optimal strokes for multi-link microswimmers: Purcell's loops and Taylor's waves reconciled}
\author{Fran{\c{c}}ois Alouges, Antonio DeSimone,  Laetitia Giraldi, \\
Yizhar Or,  and Oren Wiezel}

\maketitle

\begin{abstract}
Micron-scale swimmers move in the realm of negligible inertia, dominated by viscous drag forces. 
In this paper, we formulate the leading-order dynamics of a slender multi-link ($N$-link) microswimmer assuming small-amplitude undulations about its straight configuration. 
The energy-optimal stroke to achieve a given prescribed displacement in a given time period is obtained as the largest eigenvalue solution of a constrained optimal control problem. Remarkably, the optimal stroke is an ellipse lying within a two-dimensional plane in the ($N$-1)-dimensional space of joint angles, where $N$ can be arbitrarily large. 
For large $N$, the optimal stroke is a traveling wave of bending, modulo edge effects.

If the number of shape variables is small, we can consider the same problem when the prescribed displacement in one time period is large, and not attainable with small variations of the joint angles.
The fully nonlinear optimal control problem is solved numerically for the cases $N$=3 (Purcell's three-link swimmer) and $N$=5 showing that, as the prescribed displacement becomes small, the optimal solutions obtained using the small-amplitude assumption are recovered.
We also show that, when the prescribed displacements become large, the picture is different. For $N$=3 we recover the non-convex planar loops already known from previous studies. For $N$=5 we obtain non-planar loops, raising the question of characterizing the geometry of complex high-dimensional loops.
\end{abstract}

\section{Introduction}

The analysis of biological and bio-inspired swimming at microscopic scales has attracted considerable attention in the recent literature, starting from the seminal work by Taylor \cite{Taylor51}, Lighthill \cite{Lighthill75},  and Purcell \cite{Purcell77}. 
One of the reasons is that swimming of unicellular organisms is at the root of many fundamental processes in biology: reproduction through the swimming of sperm cells is just one example \cite{GrayHancock55,Drezner1981,Gaffney2011,Gaffney2017}. Moreover, biology inspires the design of bio-mimetic artificial devices that may have important applications in medicine as drug delivery, diagnostic or therapeutic devices (for example: smart endoscopic capsules), see e.g.  \cite{Dreyfus2005,micromachines2014,alouges2015soft}.
The large size of the recent literature makes it impossible to provide an exhaustive survey of the state of the art. Specialized reviews have appeared in recent years, 
such as \cite{Gaffney2011,LaugaPowers09,StokerReview,GoldsteinReview}, and several monographs are available \cite{Lighthill75,childress.book,Bonnard_book}. The interested reader may find in these works and in the references cited therein several hundred papers to explore the subject.

At the scale of a single cell, viscous forces dominate inertia in fluid flows, which are then governed by the (steady) Stokes equations  \cite{BrennerHappel65}. These arise from the Navier-Stokes equation, in the (formal) limit of zero Reynolds and Womersley numbers. The linearity of the Stokes system has the consequence that  net propulsion through periodic shape actuation is only possible through histories of shape changes that are not reciprocal in time, a fact popularized as ``the scallop theorem'' by Purcell \cite{Purcell77}.

Non-reciprocal periodic shape changes can arise in a number of ways: thanks to a non-trivial topology of the space of shapes, through non-trivial loops in the space of shapes, thanks to the propagation of travelling waves of deformation. Corresponding biological examples are, respectively, the rotary motion of helical bacterial flagellar bundles,
the different shapes of cilia in the power and in the recovery part of one stroke, the beating of a eukaryotic flagellum causing the propagation of bending waves along the length of the flagellum.
In fact, bending waves propagating along cilia/flagella and shape modulation during one stroke are used not only for propulsion by micro-organisms, but also for transport  inside organs in humans and other higher organisms \cite{blake_1971,Satir1990,Cicuta2017}.
The second example is the one with more connections with the study of minimal artificial swimmers (i.e., swimmers with only two internal degrees of freedom to control shape such as
the three-link swimmer of Purcell \cite{Purcell77,becker2003self}, the three-sphere swimmer of Najafi and Golestanian \cite{najafi2004simple}, and others.)
The third example is possibly the most thoroughly exploited paradigm in the fabrication of micro-swimmer prototypes, often through the action of an external magnetic field on a flexible filament \cite{Dreyfus2005,alouges2015soft,Fischer2009,Sitti2018}.

Patterns of optimal actuation have been investigated independently, for each of these three examples, with a variety of analytical  and numerical methods. For the case of flagellar and ciliary propulsion, these  include
 \cite{Lighthill75,Pironneau74,berman2013undulatory,Lauga2014,HosoiOptimalBiflagellated}  leading to recognizing, for example, helical shapes as the optimal ones for filaments in three dimensions, and smoothed saw-tooth traveling waves as the optimal wave forms for the planar beating of a one-dimensional flagellum or for a planar sheet. In the limit of small amplitudes, the latter reduces to Taylor's traveling sine waves \cite{Taylor51}. For the study of optimal strokes for minimal artificial swimmers the reader is referred, e.g., to 
 \cite{becker2003self,TamHosoi2007,alouges2013optimally,alouges2008,alougesM3AS,oren_cdc2016,Giuliani2018}.

More recently, it has been recognized that, in the presence of external forces or torques, the scallop paradigm has to be reconsidered. This is a consequence of the fact that, in this case, the governing equations can no longer be cast as an affine control systems without drift. Similar remarks apply to the presence of parts of the swimmer body whose shape is not directly controlled and it instead emerges from the balance between elastic restoring forces and viscous resistances.
Since here we will not  pursue this issue any further, we refer the reader to some of the relevant literature  \cite{Hosoi2link}  and \cite{spagnolie2010optimal,passov2012dynamics,Montino,cicconofri2016motion,Gaffney2018}.

In spite of all the recent progress, several questions remain open for investigation as testified by the growing pace at which research articles on low Reynolds number swimmers are being published. New aspects of the very same fundamental swimming problem are thus continuously emerging.
Rather than adding one more, 
we are motivated here by the quest for unifying perspectives over this vast literature.
For example: what is the connection, if any,  between optimal actuation by traveling waves inspired by Taylor's swimming sheet \cite{Taylor51} and optimal actuation by closed loops in the space of shapes shown in Tam and Hosoi's gaits \cite{TamHosoi2007} for Purcell's swimmer and further discussed in \cite{RazAvron}? The two main paradigms for producing non-reciprocal shape changes have remained mostly disconnected in the recent literature, confined to two independent streams of research. 

Motivated by this question, we focus on one specific example, namely, a planar swimmer consisting of $N$ equal links of fixed length $L$ representing the $N$-dimensional generalization of Purcell's famous three-link swimmer, and use geometric control theory 
to identify some unifying principles.
In particular, we consider the problem of prescribing an admissible displacement in one shape cycle, and look for the gait that minimizes a suitable cost functional giving a measure of the expended energy during that cycle.

The problem of optimal control is non-linear in the shape parameters, and determining (even numerically) the optimal gait explicitly becomes very hard  as soon as the number of shape parameters becomes large.
Under the assumption of small-amplitude joint angles, by considering the approximation at leading order in the shape parameters, we obtain an affine control system which can be analyzed in full detail. 
In previous studies, asymptotic analysis in this small-angle regime has been used for  Purcell's and other swimmers where the number $N$ of shape degrees of freedom is small \cite{giraldi2015optimal,wiezel2016optimization,alouges2016parking},  but never for large $N$.
We find that optimal gaits are always two-dimensional elliptical loops, independent of $N$. These gaits bridge Purcell's loops for the two-dimensional shape space associated with $N$=3, to gaits that, modulo edge effects, can be identified with Taylor's traveling waves of bending for large $N$.
Interestingly, this analysis can be done without any explicit reference to a concrete model for the interaction between the swimmer and the surrounding fluid, which can be modeled with the full detail of Stokes hydrodynamics, or with any of the simplified models to treat the viscous  drag for a slender swimmer. The result only depends on structural properties and symmetries of the governing equations, which in turn reflect the geometric symmetry of the physical problems. In fact, in this regime of small-amplitude joint angles, the perturbations from  the rectilinear geometry of the reference configuration are small, and a slender  one-dimensional swimmer with homogenous geometric and mechanical properties that interacts with a homogeneous surrounding medium is a system which is essentially invariant under shifts along the body axis. This is exactly true for an infinite or a periodic system and approximately true, modulo edge effects, for a  system of finite length. The relevance of traveling waves as optimal gaits is therefore naturally suggested by the geometric symmetry of the system and, in fact, it emerges naturally from the symmetries of the governing equations.

Our result can be seen as a unifying principle in the sense that many different concrete problems, where different geometries of micro-swimmers obtained as $N$ is varied, and different models for the fluid-structure interactions are employed, lead to different governing equations, but their behavior can be interpreted on the basis of the same single principle, thanks to the fact that  the governing equations all share the same abstract structure and symmetries.
To reinforce this point even further, we note that a similar result has been obtained in the different but related context of crawling  of one-dimensional objects on solid surfaces or within a solid matrix \cite{Agostinelli}, where it is shown that peristaltic waves used by earthworms for locomotion can be interpreted as optimal gaits when only small deformations are allowed. This is true provided, of course, that both mechanical properties of the crawler and frictional interactions with the surroundings are invariant with respect to shifts along the body axis.

To remove the assumption of small-amplitude joint angles and analyze the case in which large displacements are prescribed (and large  joint angle variations are allowed)
we can only resort to numerics, and our current capabilities limit the size of the system we can handle (in this paper, $N$=5).
This is nevertheless sufficient to reveal that optimal gaits of large amplitude have interesting and nontrivial geometry. They may differ both qualitatively and quantitatively from the planar elliptical loops representing their small amplitude limits. In particular, we find non-convex planar loops for $N$=3, as it was already well-known from the analysis of optimal gaits for  Purcell's swimmer, but also non-planar closed space curves for $N$=5.  However, all these nontrivial gaits duly converge to planar elliptical loops  when the size of the prescribed displacement in one cycle becomes small.

Seen from the perspective of our results on the small-amplitude angle regime, these results are not surprising.
In particular, when the restricted setting of small perturbations from the rectilinear geometry is abandoned, and large shape changes are considered, then invariance under shifts along the body axis is lost and traveling waves are no longer a natural basis for the study of the properties of optimal gaits.
In addition, the optimal loops are no longer planar: the study of their geometric properties emerges in this way as a new and completely open field for future investigations that will require the development of new theoretical and numerical tools.

The rest of the paper is organized as follows.
We describe the mathematical setting of the optimal control problem for the $N$-link swimmer in Section~\ref{sec:dynamics}. In Section~\ref{sec:small_amplitude} we consider the regime of small-amplitude joint angles, hence of small deviations from the rectilinear shape, derive the general result that optimal loops are planar ellipses, and we solve numerically some specific examples which, for large $N$, show the emergence of traveling waves as optimal gaits. In Section \ref{sec:large_amplitude} we move to the  regime of large amplitude joint angles, hence of large deviations from the rectilinear shape, and discuss some numerical results obtained via direct numerical optimization using the software {\sc{Bocop}} \cite{bonnans2012Bocop}: we exhibit non-planar complex loops for the case $N$=5 and show that they converge to planar ellipses as the size of the prescribed displacement in one cycle becomes small.

\section{Dynamics of the $N$-link swimmer}\label{sec:dynamics}

We develop our  analysis at leading-order of the dynamics of the $N$-link neutrally buoyant swimmer  using only the assumptions following from the linearity of Stokes equations, their invariance properties, and the symmetries of the swimmer.
These assumptions are valid for several classic models such as Purcell's swimmer \cite{Purcell77},
the $N$-link swimmer in \cite{AlougesDeSimone13}, and Taylor's swimming sheet \cite{Taylor51}. They also hold irrespectively of whether the forces exerted by the fluid surrounding the swimmer are calculated by  solving explicitly the Stokes equations, or they are evaluated by using any of the several approximate methods that have been used for slender swimmers (resistive force theory \cite{GrayHancock55}, slender body theory \cite{cox1970} etc.).

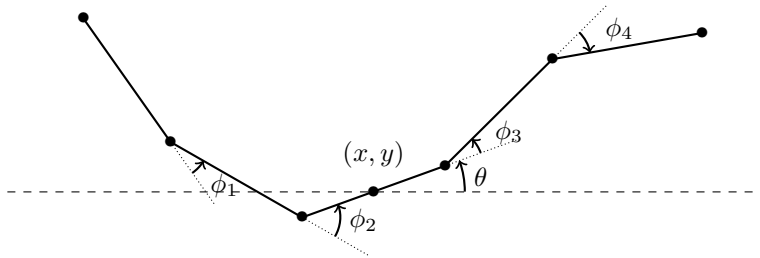
\begin{figure}
\centering
\begin{tikzpicture}
\draw[dashed] (-5,0) -- (5,0) ;
\draw [thick, mark=ball, mark size=3pt](-4,2.3)  node {$\bullet$} -- ++(-55:2)  coordinate (A) node {$\bullet$}  ;
\draw[densely dotted] (A) -- ++(-55:1) ;
\draw[thick] (A) ++(-55:0.5)[->] arc(-55:-30:0.5);
 \draw (A) ++(-55:0.7) node[right]{$\phi_1$};
 
\draw [thick, mark=ball, mark size=3pt](A)  -- ++(-30:2)  coordinate (B) node {$\bullet$}  ;
\draw[densely dotted] (B) -- ++(-30:1) ;
\draw[thick] (B) ++(-30:0.5)[->] arc(-30:20:0.5);
\draw (B) ++(-5:0.5) node[right]{$\phi_2$};
 
\draw [thick, mark=ball, mark size=3pt](B)  -- ++(20:2)  coordinate (C) node {$\bullet$}  ;
\draw[densely dotted] (C) -- ++(20:1) ;
\draw[thick] (C) ++(20:0.5)[->] arc(20:45:0.5);
\draw (C) ++(40:0.7) node[right]{$\phi_3$};
 
\draw [thick, mark=ball, mark size=3pt](C)  -- ++(45:2)  coordinate (D) node {$\bullet$}  ;
\draw[densely dotted] (D) -- ++(45:1) ;
\draw[thick] (D) ++(45:0.5)[->] arc(45:10:0.5);
\draw (D) ++(35:0.7) node[right]{$\phi_4$};

\draw [thick, mark=ball, mark size=3pt](D)  -- ++(10:2)  coordinate (D) node {$\bullet$}  ;
 
\draw[thick, mark=ball, mark size=3pt](B)++(20:1)  coordinate (X) node {$\bullet$}  ;
\draw[thick, ->](X)++(1.2,0) arc(0:20:1.2);
\draw(X)++(1.2,0.2) node[right]{$\theta$} ;
\draw(X)++(0,0.2) node[above]{$(x,y)$};
\end{tikzpicture}
 \caption{the $N$-link swimmer. Represented here with $N=5$ links. The position $(x,y)$ corresponds to the midpoint of the central link, and $\theta$ is the angle that this link makes with the horizontal axis. The shape parameters $(\phi_1,\cdots,\phi_{N-1})$ are the angles between two neighboring links at each joint.}
 \label{Fig:N-link_notation}
 \end{figure}

The swimmer is planar and composed of $N$ identical segments connected by rotational joints, see Fig \ref{Fig:N-link_notation}. We assume in the following that $N$ is odd, and denote by $(x,y)$ the position of the midpoint of the central link and $\theta$ its orientation (the angle that the central link makes with the horizontal axis), see Fig \ref{Fig:N-link_notation}.
The case of even $N$ simply requires a different choice of coordinates.
The swimmer is immersed in an unbounded domain occupied by a homogeneous viscous fluid governed by Stokes equations and is driven by shape changes, i.e., the joint angles $\vecphi=(\phi_1,\phi_2,\ldots\phi_{N-1})^T$ are given functions of time $t$. In particular, we will be concerned with $T-$periodic shape changes $\vecphi(t)$, which we call strokes, satisfying $\vecphi(t) = \vecphi(T+t), \quad \forall t>0$. Our analysis aims at resolving the leading-order terms of the swimmer's dynamics, and is restricted to up to second order $O(\vep^2)$ terms in the shape parameters $\phi_i(t)=O(\vep)$, which are assumed to be small, i.e. $\vep \ll 1$.

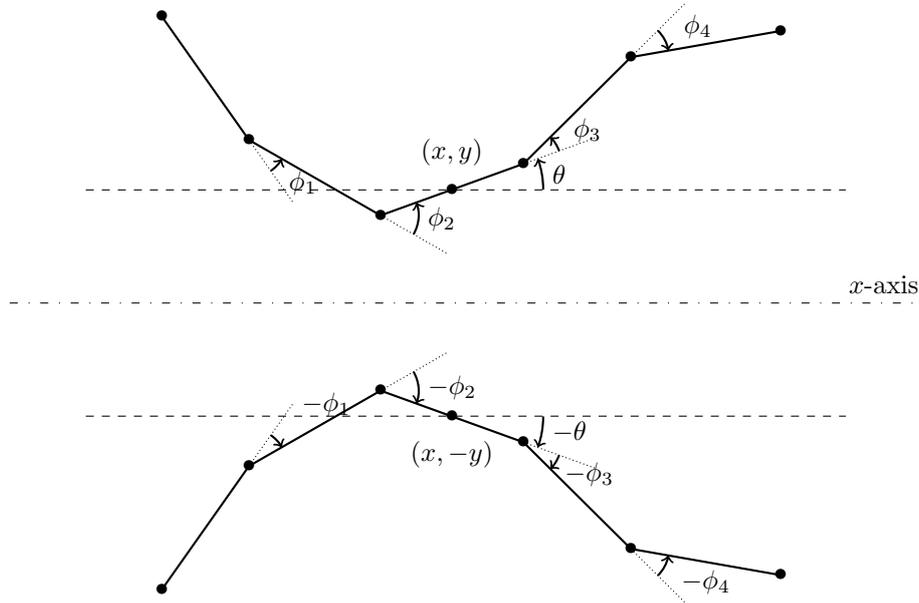
\begin{figure}[bht]
\centering
\begin{tikzpicture}
\draw[dashed] (-5,0) -- (5,0) ;
\draw [thick, mark=ball, mark size=3pt](-4,2.3)  node {$\bullet$} -- ++(-55:2)  coordinate (A) node {$\bullet$}  ;
\draw[densely dotted] (A) -- ++(-55:1) ;
\draw[thick] (A) ++(-55:0.5)[->] arc(-55:-30:0.5);
\draw (A) ++(-55:0.7) node[right]{$\phi_1$};
 
\draw [thick, mark=ball, mark size=3pt](A)  -- ++(-30:2)  coordinate (B) node {$\bullet$}  ;
\draw[densely dotted] (B) -- ++(-30:1) ;
\draw[thick] (B) ++(-30:0.5)[->] arc(-30:20:0.5);
\draw (B) ++(-5:0.5) node[right]{$\phi_2$};
 
\draw [thick, mark=ball, mark size=3pt](B)  -- ++(20:2)  coordinate (C) node {$\bullet$}  ;
\draw[densely dotted] (C) -- ++(20:1) ;
\draw[thick] (C) ++(20:0.5)[->] arc(20:45:0.5);
\draw (C) ++(40:0.7) node[right]{$\phi_3$};
 
\draw [thick, mark=ball, mark size=3pt](C)  -- ++(45:2)  coordinate (D) node {$\bullet$}  ;
\draw[densely dotted] (D) -- ++(45:1) ;
\draw[thick] (D) ++(45:0.5)[->] arc(45:10:0.5);
\draw (D) ++(35:0.7) node[right]{$\phi_4$};

\draw [thick, mark=ball, mark size=3pt](D)  -- ++(10:2)  coordinate (D) node {$\bullet$}  ;
 
\draw[thick, mark=ball, mark size=3pt](B)++(20:1)  coordinate (X) node {$\bullet$}  ;
\draw[thick, ->](X)++(1.2,0) arc(0:20:1.2);
\draw(X)++(1.2,0.2) node[right]{$\theta$} ;
\draw(X)++(0,0.2) node[above]{$(x,y)$};

\draw[loosely dashdotted](-6,-1.5) -- (6,-1.5) ; 
\draw(5.5,-1.5) node[above]{$x$-axis};
\draw[dashed] (-5,-3) -- (5,-3) ;
\draw [thick, mark=ball, mark size=3pt](-4,-5.3)  node {$\bullet$} -- ++(55:2)  coordinate (A) node {$\bullet$}  ;
\draw[densely dotted] (A) -- ++(55:1) ;
\draw[thick] (A) ++(55:0.5)[->] arc(55:30:0.5);
\draw (A) ++(55:1) node[right]{$-\phi_1$};
 
\draw [thick, mark=ball, mark size=3pt](A)  -- ++(30:2)  coordinate (B) node {$\bullet$}  ;
\draw[densely dotted] (B) -- ++(30:1) ;
\draw[thick] (B) ++(30:0.5)[->] arc(30:-20:0.5);
\draw (B) ++(5:0.5) node[right]{$-\phi_2$};
 
\draw [thick, mark=ball, mark size=3pt](B)  -- ++(-20:2)  coordinate (C) node {$\bullet$}  ;
\draw[densely dotted] (C) -- ++(-20:1) ;
\draw[thick] (C) ++(-20:0.5)[->] arc(-20:-45:0.5);
\draw (C) ++(-45:0.6) node[right]{$-\phi_3$};
 
\draw [thick, mark=ball, mark size=3pt](C)  -- ++(-45:2)  coordinate (D) node {$\bullet$}  ;
\draw[densely dotted] (D) -- ++(-45:1) ;
\draw[thick] (D) ++(-45:0.5)[->] arc(-45:-10:0.5);
\draw (D) ++(-39:0.7) node[right]{$-\phi_4$};

\draw [thick, mark=ball, mark size=3pt](D)  -- ++(-10:2)  coordinate (D) node {$\bullet$}  ;
 
\draw[thick, mark=ball, mark size=3pt](B)++(-20:1)  coordinate (X) node {$\bullet$}  ;
\draw[thick, ->](X)++(1.2,0) arc(0:-20:1.2);
\draw(X)++(1.2,-0.2) node[right]{$-\theta$} ;
\draw(X)++(0,-0.8) node[above]{$(x,-y)$};
\end{tikzpicture}
\caption{Symmetry with respect to the $x$-axis. The swimmer at position $(x,y,\theta)$ and with shape $(\phi_1,\cdots,\phi_{N-1})$ is symmetrized into a swimmer at position $(x,-y,-\theta)$ with shape $(-\phi_1,\cdots,-\phi_{N-1})$.}
\label{Fig:symmetry_figs} 
\end{figure}

The dynamics of the swimmer's planar motion is governed by
\begin{equation} \label{eq.dyn}
\left\{
\begin{array}{l}
\dot{x} = \vecf(\theta,\vecphi)\cdot \dot{\vecphi}\,,\\
\dot{y} = \vecg(\theta,\vecphi)\cdot \dot{\vecphi}\,,\\
\dot{\theta} = \vech(\vecphi)\cdot \dot{\vecphi}\,.\\
\end{array}
\right.
\end{equation}
These equations come from the balance of viscous force and torque, which are linear in $\dot{x},\dot{y},\dot{\theta}$ and $\dot{\vecphi}$. Details of the derivation in the case of  resistive force theory \cite{GrayHancock55} are given in \cite{wiezel2016optimization,AlougesDeSimone13,giraldi2013controllability}. The special structure of \eqref{eq.dyn}, namely the fact that functions $\vecf$, $\vecg$, and $\vech$  are independent of $x$ and $y$, and that the last one is also independent of $\theta$, are consequences of the translational and rotational invariance of the problem.

Due to the geometric structure of the swimmer, the functions $\vecf,\vecg$ and $\vech$ that define the dynamics satisfy further relations that are deduced from symmetries of the system. Among them, the symmetry with respect to the $x-$axis,  depicted in Fig. \ref{Fig:symmetry_figs}, transforms the swimmer parametrized by $\vecphi$ at position $(x,y,\theta)$ to the one parametrized by $-\vecphi$, at position $(x,-y,-\theta)$. The invariance of the dynamics under such a transformation leads to
\begin{equation}
\left\{
\begin{array}{l}
\vecf(\theta,\vecphi)\cdot \dot{\vecphi} = \vecf(-\theta,-\vecphi)\cdot (-\dot{\vecphi})\,,\\
\vecg(\theta,\vecphi)\cdot \dot{\vecphi} = -\vecg(-\theta,-\vecphi)\cdot (-\dot{\vecphi})\,,\\
\vech(\vecphi)\cdot \dot{\vecphi} = -\vech(-\vecphi)\cdot (-\dot{\vecphi})\,.
\end{array}
\right.
\label{symmx}
\end{equation}

\subsection{Dynamics at leading order}
\label{subsec:leading_order_dynamics}
For $\theta$ and $\vecphi$ of order $\vep$, we expand the dynamics \eqref{eq.dyn} to second order in $\theta$, $\vecphi$ and $\dot{\vecphi}$ by expanding $\vecf$ to first order as
\begin{equation}
\vecf(\theta,\vecphi) \sim \vecF_0 + \vecF_\theta \theta + \vecF_{\vecphi} \vecphi+O(\vep^2)
\label{expansion}
\end{equation}
where $\vecF_0=\vecf(0,0)\in \mathbb{R}^{N-1}$, $\vecF_\theta=\frac{\partial \vecf}{\partial \theta}(0,0)\in \mathbb{R}^{N-1}$ and $\vecF_\phi=\frac{\partial \vecf}{\partial \vecphi}(0,0)\in \mathbb{R}^{N-1,N-1}$. We also expand similarly $\vecg$ and $\vech$.
Using the symmetry relations \eqref{symmx}, we get for $\vecf$, $\vecg$ and $\vech$ respectively
\begin{eqnarray*}
\left(\vecF_0 + \vecF_\theta \theta + \vecF_{\vecphi} \vecphi\right) \cdot \dot{\vecphi} &=& -\left(\vecF_0 -  \vecF_\theta \theta - \vecF_{\vecphi} \vecphi\right) \cdot \dot{\vecphi}\\
\left(\vecG_0 + \vecG_\theta \theta + \vecG_{\vecphi} \vecphi\right) \cdot \dot{\vecphi} &=& \left(\vecG_0 -  \vecG_\theta \theta - \vecG_{\vecphi} \vecphi\right) \cdot \dot{\vecphi}\\
\left(\vecH_0 + \vecH_{\vecphi} \vecphi\right) \cdot \dot{\vecphi} &=& \left(\vecH_0 - \vecH_{\vecphi} \vecphi\right) \cdot \dot{\vecphi},
\end{eqnarray*}
from which we deduce, $\vecphi$ and $\dot{\vecphi}$ being arbitrary, that $\vecF_0=\vecG_\theta = 0$ and 
$\vecG_{\vecphi}=\vecH_{\vecphi}=0$. The original system \eqref{eq.dyn} therefore reduces to
\begin{equation} \label{eq.dyn2}
\left\{
\begin{array}{l}
\dot{x} = \left(\vecF_\theta \theta + \vecF_{\vecphi} \vecphi\right)\cdot \dot{\vecphi}+O(\varepsilon^3)\,,\\
\dot{y} = \vecG_0\cdot \dot{\vecphi}+O(\varepsilon^3)\,,\\
\dot{\theta} = \vecH_0\cdot \dot{\vecphi}+O(\varepsilon^3)\,.\\
\end{array}
\right.
\end{equation}

Integrating in time we obtain
\begin{equation}
\theta(t) = \int_0^t \dot{\theta} \,dt = \int_0^t\vecH_0\cdot \dot{\vecphi}\,dt =\vecH_0\cdot (\vecphi(t)-\vecphi(0))+O(\vep^3)
\label{eqtheta}
\end{equation}
and
$$
y(t) = \int_0^t \dot{y} \,dt = \int_0^t\vecG_0\cdot \dot{\vecphi}\,dt =
\vecG_0 \cdot (\vecphi(t)-\vecphi(0))+O(\vep^3) \,.
$$

We deduce that, up to second order, the swimmer experiences no global rotation or transverse translation after one complete stroke. Indeed, $\vecphi$ being $T$-periodic, we have
\begin{equation}
\Delta \theta = \theta(T) - \theta(0) = O(\varepsilon^3)\,,\quad\quad \Delta y = y(T) - y(0) = O(\varepsilon^3)\,.\label{eq:delta_th_y}
\end{equation}

Moreover, using the expression for $\theta$ in \eqref{eqtheta} in the first equation of \eqref{eq.dyn2} and
integrating over one period leads to 
$$
\Delta x = x(T)-x(0) =  \int_0^T \left(\vecH_0^T  \otimes \vecF_{\theta} + \vecF_{\vecphi} \right)\vecphi \cdot \dot{\vecphi}\,dt + O(\varepsilon^3) \,.
$$
Calling $\vecF^{\theta}_{\vecphi}$ the matrix $\vecH_0^T  \otimes \vecF_{\theta} + \vecF_{\vecphi}$, integrating by parts the last expression, and using again the $T$-periodicity of $\vecphi$ leads to
$$
\int_0^T \vecF^{\theta}_{\vecphi} \vecphi \cdot \dot{\vecphi}\,dt = - \int_0^T \vecF^{\theta}_{\vecphi} \dot{\vecphi} \cdot \vecphi\,dt = - \int_0^T {\vecF^{\theta}_{\vecphi}}^T \vecphi \cdot \dot{\vecphi}\,dt  \,.
$$
Finally we get the net longitudinal displacement in one cycle as
\begin{align}
\Delta x
 &= \int_0^T   \frac12 (\vecF^{\theta}_{\vecphi}-{\vecF^{\theta}_{\vecphi}}^T)  \vecphi \cdot \dot{\vecphi}\,dt+O(\vep^3) \label{eq:deltax}\,.
\end{align}

\begin{remark}
Our analysis at order two in the joint-angle amplitudes proves that the net lateral displacement $\Delta y$ and the net rotation both vanish at order two, as given in Eq. \eqref{eq:delta_th_y}, while $\Delta x$ may be non-zero at order two, as shown in Eq. \eqref{eq:deltax}. The fundamental physical explanation of this result lies in the fact that the swimmer has symmetry about its straightened configurations, and only time-periodic strokes consisting of small-amplitude deviations about this configuration are analyzed. These are zero-mean trajectories, which, due to the swimmer’s symmetry, cancel out all displacements at leading order, except for the X displacement which is along the swimmer’s axis of symmetry. This fact has also been previously explained in some related works \cite{Avron2008cousin,Emiliya2016symmetries&gaits,alouges2016parking}. This is also related to the fact the Lie-bracket vector field for each pair of the joint angle inputs, evaluated at the zero (straight) configuration, gives only $x-$displacement while all other motions only appear in higher order Lie brackets, see \cite{probprog96}.
\end{remark}

\begin{remark}
Other symmetries may be also used to obtain further information about the remaining coefficients of the system. For instance, using the rotational invariance of the system one can show that
$$
\left\{
\begin{array}{l}
\vecf(\theta,\vecphi) = \cos(\theta)\vecf(0,\vecphi) -\sin(\theta) \vecg(0,\vecphi)\,,\\
\vecg(\theta,\vecphi) = \sin(\theta)\vecf(0,\vecphi) +\cos(\theta) \vecg(0,\vecphi)\,,
\end{array}
\right.
$$
from which one easily obtains $\vecF_\theta = -\vecG_0$ (and $\vecG_\theta = \vecF_0 = 0$). Similarly, using the symmetry of the system with respect to the $y$-axis, it can be shown that both $\vecF_\theta$ and $\vecG_0$ are `even' vectors (in the sense that $\vecF_{\theta,i}=\vecF_{\theta,N-i}$ and similarly for $\vecG_0$) while $\vecH_0$ is an `odd' vector (for which $\vecH_{0,i}=-\vecH_{0,N-i}$\,.)
\end{remark}

\subsection{Power expended}
We now derive a leading-order expression for the mechanical energy expended by the swimmer during one cycle. The instantaneous power (i.e. rate of mechanical work) is given by the scalar product between force and velocity densities integrated over its surface (see e.g. \cite{Lighthill75}).
Moreover, due to the linearity of Stokes equations, both the forces and the velocities acting on the swimmer depend linearly on $\vecphid$.
Thus, the instantaneous power density expended by the swimmer is a quadratic form in $\vecphid$, with coefficients depending on $\vecphi$ (see \cite{wiezel2016optimization} for concrete expressions using resistive force theory). We may therefore write the total energy $E$ expended by the swimmer during the stroke as
\begin{equation}
\label{eq:energy_full}
E(\vecphi) = \int_0^T < \vecP(\vecphi) \vecphid,\vecphid>\,dt \,,
\end{equation}
where $\vecP(\vecphi)\in \mathcal{M}_{N,N}(\mathbb{R})$ is a symmetric and positive definite matrix, and the bracket $<\cdot,\cdot>$ stands for the scalar product in $\mathbb{R}^N$.
Expanding this expression gives the $O(\vep^2)$ leading-order term of the energy as
\begin{equation}
\label{eq:energy}
E_0(\vecphi) = \int_0^T < \vecP_0 \vecphid,\vecphid>\,dt \,,
\end{equation}
where $\vecP_0 := \vecP(\vecphi=0)$ is symmetric and positive definite.

\section{Optimal strokes of small amplitude}\label{sec:small_amplitude}

In this section, we study energy-optimal strokes while focusing on the case in which the joint angles remain small (small amplitude approximation), so that only small perturbations of the rectilinear geometry are allowed.
\subsection{Analysis of the optimal control problem}

We begin by reviewing the well-known criterion of optimal energy efficiency due to Lighthill, and discuss its relation with the energy-optimal strokes studied here. Lighthill's efficiency of swimming is defined as the ratio of $E_{drag} / E_{swim}$, where $E_{drag}$ is the energy needed by an external actor to pull the swimmer during a time $T$ at an average speed $\Delta x/T$, while $E_{swim}$ is the energy expended by the swimmer during a stroke to propel itself at the same average speed.
As the drag force is proportional to the velocity, $E_{drag}$ is proportional to $(\Delta x)^2/T$ whereas $E_{swim}$ behaves like $\Delta x$ as \eqref{eq:deltax} and \eqref{eq:energy_full} indicate ($\Delta x$ and $E$ being quadratic in $(\vecphi,\dot{\vecphi})$
are of the same order $O(\varepsilon^2)$). Thus,  Lighthill's efficiency is proportional to $\Delta x$, and maximizing it amounts to increase $\Delta x$, and thus the amplitude of the stroke as much as possible, beyond the range of small angles.
This implies that optimal strokes that maximize Lighthill's efficiency typically involve large-angle trajectories \cite{TamHosoi2007,wiezel2016optimization}.
On the other hand, if the amplitude of the stroke is constrained, by e.g. the maximum allowed amplitude of the angles, the displacement will be constrained as well, and it is expected that this constraint will be saturated when attempting to maximize Lighthill's efficiency, making the resulting optimal strokes of limited general interest.
We therefore turn our attention to the problem of maximizing the efficiency (i.e. minimizing the energy) for a displacement achieved during one stroke which is bounded, or more simply fixed. Optimal strokes are thus defined as $T$-periodic strokes $\vecphi(t)$ that expend the minimal energy $E$, among all strokes that achieve a given displacement $\Delta x$ in a given time period $T$.

We showed in the previous section that, at leading order, the $y$ and $\theta$ displacements are negligible with respect to the $x$ displacement. Optimal strokes that provide a (longitudinal) displacement $\Delta x$ are thus sought as solutions to the constrained minimization problem
\begin{equation}
\label{eq:problem}
\min_{\vecphi \in \mathcal{D}} E_0(\vecphi)\,,
\end{equation}
where, using \eqref{eq:deltax}, we have
$$
\mathcal{D}:=\left\{\vecphi \in (\mathcal{C}^1(0,T))^{N-1}  | \vecphi(0) = \vecphi(T)\,\textrm{and}\,\frac12\int_0^T\!\! (\vecF^{\theta}_{\vecphi}-{\vecF^{\theta}_{\vecphi}}^T)  \vecphi \cdot \dot{\vecphi}dt = \Delta x\right\}\!.
$$

In the following, we show that the solution of this optimization problem describes a planar ellipse in  shape space, which is $(N-1)$-dimensional. We also give a method for its computation. Note that discrepancies between the solutions of the optimal control problem defined above under the leading-order approximation and under the exact nonlinear dynamics in \eqref{eq.dyn} and \eqref{eq:energy_full} become smaller and smaller by decreasing the displacement $\Delta x$ and making the stroke amplitude $\vep$ of order $O(\sqrt{\Delta x})$.\\

In order to find the Euler-Lagrange first-order necessary condition associated with the constrained optimization problem (\ref{eq:problem}), we set 
$$
\tilde{\mathcal{E}}_0(\vecphi)= \mathcal{E}_0(\vecphi)+\lambda K_0(\vecphi)
$$
where $\displaystyle K_0(\vecphi) =  \frac12 \int_0^T (\vecF^{\theta}_{\vecphi}-{\vecF^{\theta}_{\vecphi}}^T)  \vecphi \cdot \dot{\vecphi}\,dt$ and $\lambda \in \mathbb{R}$ is the Lagrange multiplier associated with the constraint $K_0(\vecphi) = \Delta x$.
Writing that the first order functional derivative of $\tilde{\mathcal{E}}_0$ vanishes at the optimal stroke $\vecphi^*$
$$
\delta \tilde{\mathcal{E}}_0(\vecphi^*)=0
$$
amounts to writing 
\begin{equation}
\frac{d}{dt}\frac{\partial L}{\partial \dot{\vecphi}}(\vecphi^*,\dot{\vecphi}^*) -\frac{\partial L}{\partial \vecphi}(\vecphi^*,\dot{\vecphi}^*)=0
\label{EL}
\end{equation}
where the Lagrangian $L$ is such that 
$$
\tilde{\mathcal{E}}_0(\vecphi) = \int_0^T L(\vecphi,\dot{\vecphi})\,dt\,.
$$
Here, we have
$$
L(\vecphi,\dot{\vecphi}) = <\vecP_0\dot{\vecphi}, \dot{\vecphi}> +\frac{\lambda}{2}  (\vecF^{\theta}_{\vecphi}-{\vecF^{\theta}_{\vecphi}}^T)  \vecphi \cdot \dot{\vecphi}
$$
and therefore \eqref{EL} becomes
\begin{equation}
\label{eq:optimal_stroke}
 \vecP_0\ddot{\vecphi}^* = \frac{\lambda}{2} ({\vecF^{\theta}_{\vecphi}}^T-\vecF^{\theta}_{\vecphi})  \dot{\vecphi}^* \,.
\end{equation}

We denote by $\vecM$ the skew symmetric matrix
\begin{equation}
\label{eq:matrix_M}
\vecM = \frac12\vecP_0^{-\frac12}({\vecF^{\theta}_{\vecphi}}^T-\vecF^{\theta}_{\vecphi})\vecP_0^{-\frac12}\,,
\end{equation}
and decompose the equation \eqref{eq:optimal_stroke} along the eigen-elements of $\vecM$. Eigenvectors of skew symmetric
matrices go by pairs, associated with conjugate and purely imaginary eigenvalues.
We therefore set $(\vecv^\pm_j)_{1\leq j\leq N'}$ ($N'=\lfloor \frac{N}{2} \rfloor$) the (complex and orthonormal) eigenvectors associated with the purely imaginary eigenvalue $\pm i\mu_j$ with $\mu_j\geq 0$:
$$
\vecM\vecv^\pm_j = \pm i\mu_j \vecv^\pm_j\,.
$$
Projecting $\vecP_0^\frac12\vecphi^*$ on the $(\vecv^\pm_j)_{1\leq j\leq N'}$ as
$$
\vecP_0^\frac12\vecphi^*(t) = \sum_{j=1}^{N'} \psi^\pm_j(t)\vecv^\pm_j,
$$
we deduce from (\ref{eq:optimal_stroke}) that
$$
\ddot{\psi}^\pm_j  = \pm i \lambda \mu_j \dot{\psi}^\pm_j \,,
$$
or  $\psi^\pm_j(t) =  \frac{\alpha_j^\pm}{\pm i \lambda \mu_j} \exp(\pm i \lambda \mu_j t)+ C^\pm_j$ where $C^\pm_j$ is a constant that we may take equal to 0.
The solution of \eqref{eq:optimal_stroke} is thus expressed as

\begin{equation}
\label{eq:phi_star}
	\vecphi^*(t) = \sum_{j=1}^{N'}  \frac{ \alpha^\pm_j\, \text{e}^{\pm i \lambda \mu_j t}}{\pm i \lambda \,\mu_j}  \vecP_0^{-1/2} \vecv^\pm_j,\quad \text{for $t \in [0,T]$}\,.
\end{equation}

Since we focus on periodic strokes, $\vecphi(0)=\vecphi(T)$, we must have
\begin{equation}
\label{eq:periodic_stroke}
\lambda \mu_j T = 2 \pi k_j\quad 1\leq j \leq N', k_j \in \mathbb{N}.
\end{equation}
By plugging the solution \eqref{eq:phi_star} into \eqref{eq:energy}, we find
\begin{equation}
\label{eq:energy_0}
E_0(\vecphi^*) = T\,\sum_{j=1}^{N'} (|\alpha^+_j|^2+|\alpha^-_j|^2)\,,
\end{equation}
while the $x$-displacement is given by
\begin{equation}
\label{eq:displacement_0}
\Delta x = T\,\sum_{j=1}^{N'} \frac{(|\alpha^+_j|^2+|\alpha^-_j|^2)}{\lambda} = \frac{E_0(\vecphi^*)}{\lambda}\,.
\end{equation}

Since $\Delta x$ is fixed in the optimization problem, minimizing the energy requires choosing $\lambda$ as small as possible.
In view of \eqref{eq:periodic_stroke}, this is achieved if $\lambda=\pm\frac{2\pi}{\mu_{\scriptscriptstyle M}T}$, where $\mu_{\scriptscriptstyle M} = \max\left\{\mu_j\,,1\leq j \leq N' \right\}$, the direction of the translation depending on the sign of $\lambda$, and $\psi^\pm_j=0$ if $\mu_j\ne \mu_{\scriptscriptstyle M}$. Assuming $\lambda>0$ ($\lambda<0$ is handled similarly), we deduce that $\lambda = 2\pi/{T \mu_{\scriptscriptstyle M}}$ and the solution has only two modes corresponding to $\vecv^\pm_{\scriptscriptstyle M}$
\begin{equation}
\label{eq:phi_star_final}
	\vecphi^*(t) =  \frac{\alpha^+_{\scriptscriptstyle M}\, \text{e}^{ 2i\pi  t}}{2 i \pi}  \vecP_0^{-1/2} \vecv^+_{\scriptscriptstyle M} - \frac{\alpha^-_{\scriptscriptstyle M}\, \text{e}^{- 2i\pi  t}}{2 i \pi}  \vecP_0^{-1/2} \vecv^-_{\scriptscriptstyle M},\quad \text{for $t \in [0,T]$}\,,
\end{equation}
and is therefore an ellipse drawn in the plane $(\vecP_0^{-1/2} \vecv^+_{\scriptscriptstyle M},\vecP_0^{-1/2} \vecv^-_{\scriptscriptstyle M})$. Furthermore, noticing that $\vecphi^*$ is real and $\vecv_{\scriptscriptstyle M}^+ = \overline{\vecv_{\scriptscriptstyle M}^-}$, we must have $\alpha_{\scriptscriptstyle M}^+ = \overline{\alpha_{\scriptscriptstyle M}^-}$. The net-displacement achieved by the gait \eqref{eq:phi_star_final} will be $\Delta x = \frac{2T}{\lambda}|\alpha^+_{\scriptscriptstyle M}|^2$. Therefore, the optimal gait amplitude scales as $\sqrt{\Delta x}$, as expected from \eqref{eq:deltax}.

Let us point out the important observation that the optimal gait \eqref{eq:phi_star_final} is always  an ellipse lying within a two-dimensional plane regardless of the number of links $N$, which makes the dimension of the shape space arbitrarily large. In addition, the optimal gait in \eqref{eq:phi_star_final} can be written equivalently as sinusoidal inputs:
\begin{equation}
\label{eq:sin}
	\phi_k^*(t) =  a_k \sin(\omega t + p_k), \quad k=1,\ldots,N\!-\!1 
\end{equation}
where $\omega = 2 \pi /T$. The amplitudes $a_k$ in \eqref{eq:sin} are of order $O(\vep)$, and scale as $\sqrt{\Delta x}$. Finally, an important
property of the optimal solution $\vecphi^*(t)$ is that it satisfies
\begin{equation}
P(t)=< \vecP_0 \vecphid(t),\vecphid(t)>=2|\alpha_{\scriptscriptstyle M}^+|^2=const\,,
\label{eq:constant_power}
\end{equation} 
where $P(t)$ is the mechanical power generated by the swimmer. Equation \eqref{eq:constant_power} implies that the optimal gait generates a constant power over the entire cycle. 
A similar result for the three-sphere swimmer was proved in \cite{alouges2008} and 
this agrees with a fundamental observation made in \cite{becker2003self}, which states that for any given trajectory in shape space, the time-parametrization associated with constant power is the one that minimizes the total energy expenditure.

\subsection{Numerical results: from Purcell's loops to Taylor's waves}
\label{Sec:Numerical}

In this section, we present numerical results for the problem of computing optimal gaits for slender swimmers by using the derivation described above and leading to the Euler-Lagrange equations \eqref{EL}. In order to derive a simple and concrete formulation of the dynamics, we use here the local drag approximation of resistive force theory \cite{GrayHancock55,cox1970} for slender links.
Our computations are made using {\sc{Matlab}}. A similar approach based on the the Euler-Lagrange equations \eqref{EL} was used in \cite{alouges2008} for the three-sphere swimmer where, however,  nonlocal hydrodynamic forces were fully resolved by solving the Stokes system for the surrounding fluid.

Our numerical code is based on the computation of $\vecphi^*(t)$ from formula \eqref{eq:phi_star_final}. 
The first steps do not rely on the small amplitude approximation, and they will be used in later sections as well.

First, we derive the dynamics of the swimmer in \eqref{eq.dyn} using resistive force theory \cite{cox1970,GrayHancock55}. This theory states that the viscous drag force $\vecf_i$  and torque $m_i$  on the $i^{th}$ slender link of length $l$ under planar motion are proportional to its linear and angular velocities, respectively. Thus, one can write the expression for the drag force and torque exerted on each link:
\begin{equation} \label{eq.rft}
\begin{array}{c}
\vecf_i=-c_t l(\vecv_i\cdot\vect_i)\vect_i-c_n l(\vecv_i\cdot\vecn_i)\vecn_i \\[12pt]
m_i=-\dfrac{1}{12} c_n l^3 \omega_i,
\end{array}
\end{equation}
where $\vecv_i$, $\omega_i$ are the linear and angular velocities of the $i^{th}$ link, and $\vect_i,\; \vecn_i$  are unit vectors in its axial and normal directions. The resistance coefficients in \eqref{eq.rft} are $c_n\!=\!2c_t\!=4\pi\eta/\log(l/a)$ where $\eta$  is the dynamic viscosity of the fluid and $a\ll l$ is the radius of the slender links' cross-section. Using \eqref{eq.rft}, the swimmer's dynamic equations can be derived from force and torque balance, see for instance \cite{wiezel2016optimization} for Purcell's swimmer and \cite{AlougesDeSimone13} for the general $N$-link swimmer. 

Next, we exploit the assumption of small amplitude approximation and we calculate the matrices $\vecF^{\theta}_{\vecphi}$ and $\vecP_0$ in \eqref{eq:deltax}, \eqref{eq:energy} associated with the leading-order expressions for the dynamics and the mechanical energy expenditure. Then it only remains to compute the eigenvalues and eigenvectors of the matrix $\vecM$ in \eqref{eq:matrix_M} in order to assemble the expression of the optimal gait $\vecphi^*(t)$ in \eqref{eq:phi_star_final}. This computation is done numerically in {\sc{Matlab}}.

\subsection{Purcell's swimmer and $N$-link swimmers for small $N$}\label{sec:Purcell}

We use the small amplitude approximation and solve equation  \eqref{eq:optimal_stroke} numerically  to compute optimal  strokes for Purcell's three link swimmer and for other cases with small $N$. Purcell's three-link swimmer is the minimal model of a linked microswimmer. We used the formula \eqref{eq:phi_star_final} in order to compute the energy-optimal gait, which is shown in Fig. \ref{fig:optimal_gait_3link} as a trajectory in the plane of joint angles $(\phi_1,\phi_2)$ for $\Delta x = 0.1l$. Snapshots of the swimmer's configuration at quarter-period times are also illustrated on the plot, and we recover the key ingredient that enables the propulsion of this system, namely a phase shift between the two actuators.

The case of $N=5$ link swimmer is also computed and solutions are shown in Fig.~\ref{Fig:5linksnap}. Four snapshots of the swimmer obtained again at quarter-period times are presented corresponding to a sequence of shapes along the optimal stroke.

Both cases ($N=3$ and $N=5$) will be discussed further and compared with those obtained without assuming that the prescribed displacements are small and joint angles have small amplitude in Section~\ref{sec:large_amplitude}.

\subsection{Optimal gaits for swimmers with many links}
\begin{figure}[ht]
\centering
\begin{subfigure}{0.29\textwidth}
\includegraphics[width=\textwidth]{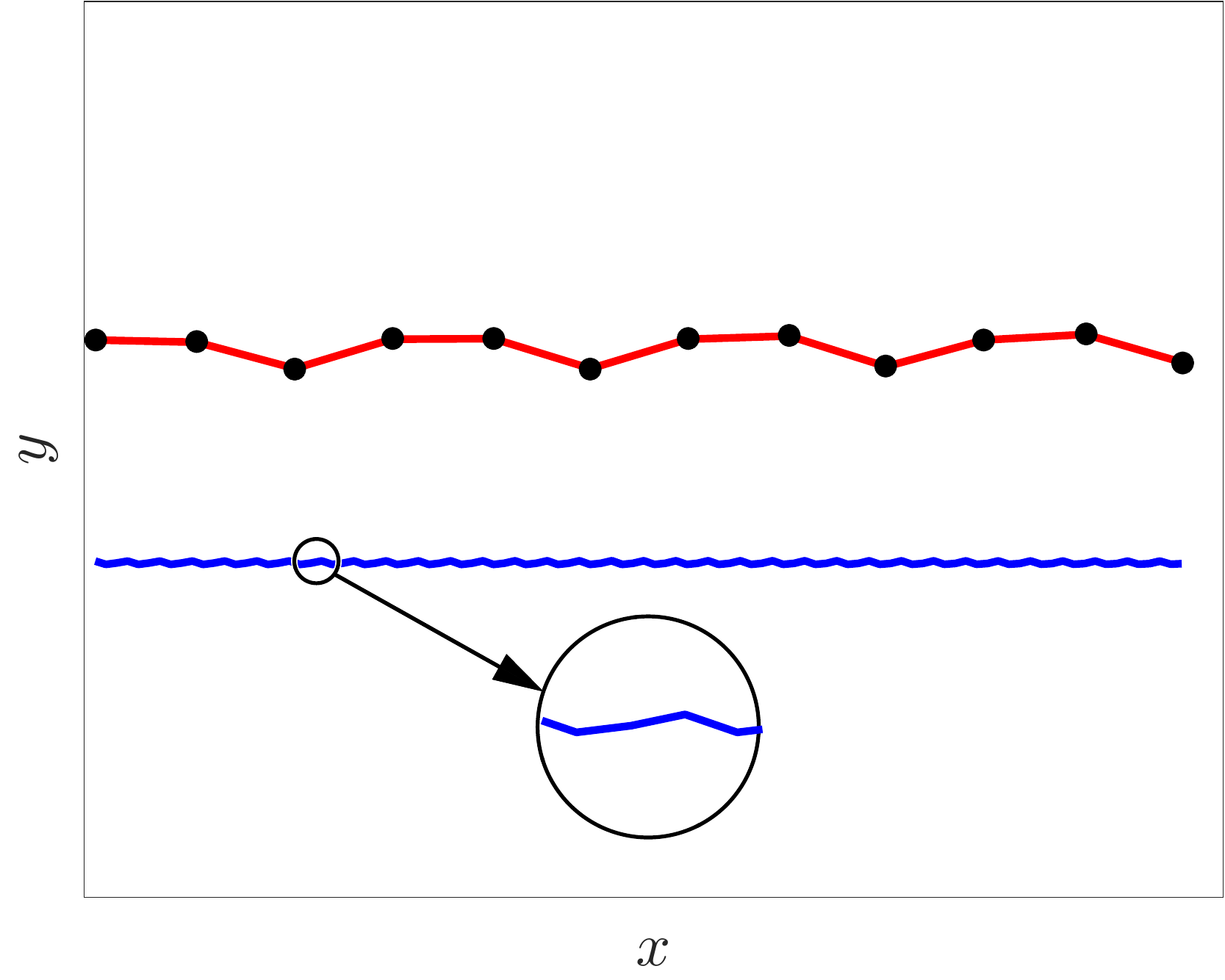}
\caption{}
\label{fig:snapshot_101&11}
\end{subfigure}
\begin{subfigure}{0.33\textwidth}
\includegraphics[width=\textwidth]{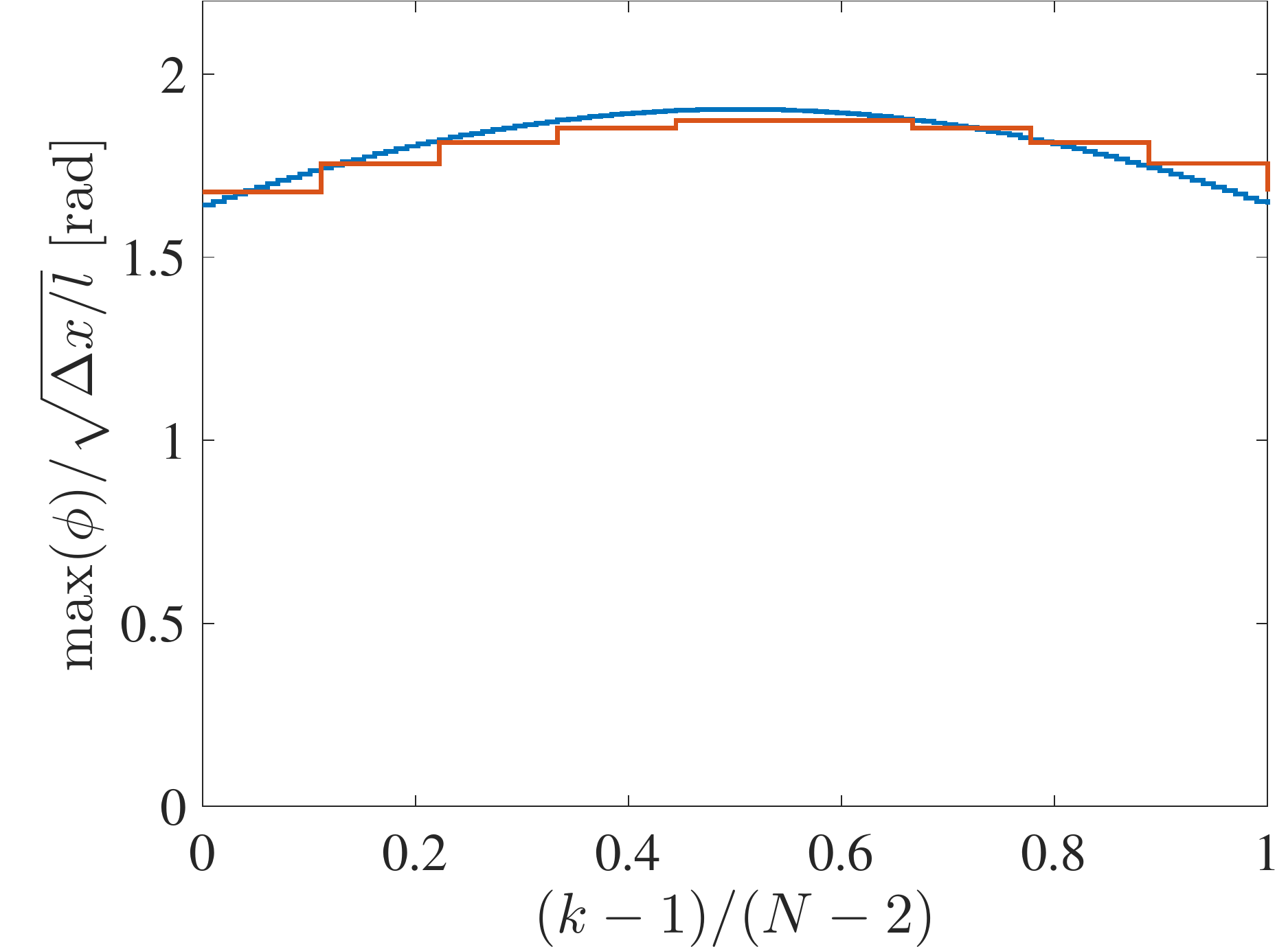}
\caption{}
\label{fig:amplitude101&11}
\end{subfigure}
\begin{subfigure}{0.33\textwidth}
\includegraphics[width=\textwidth]{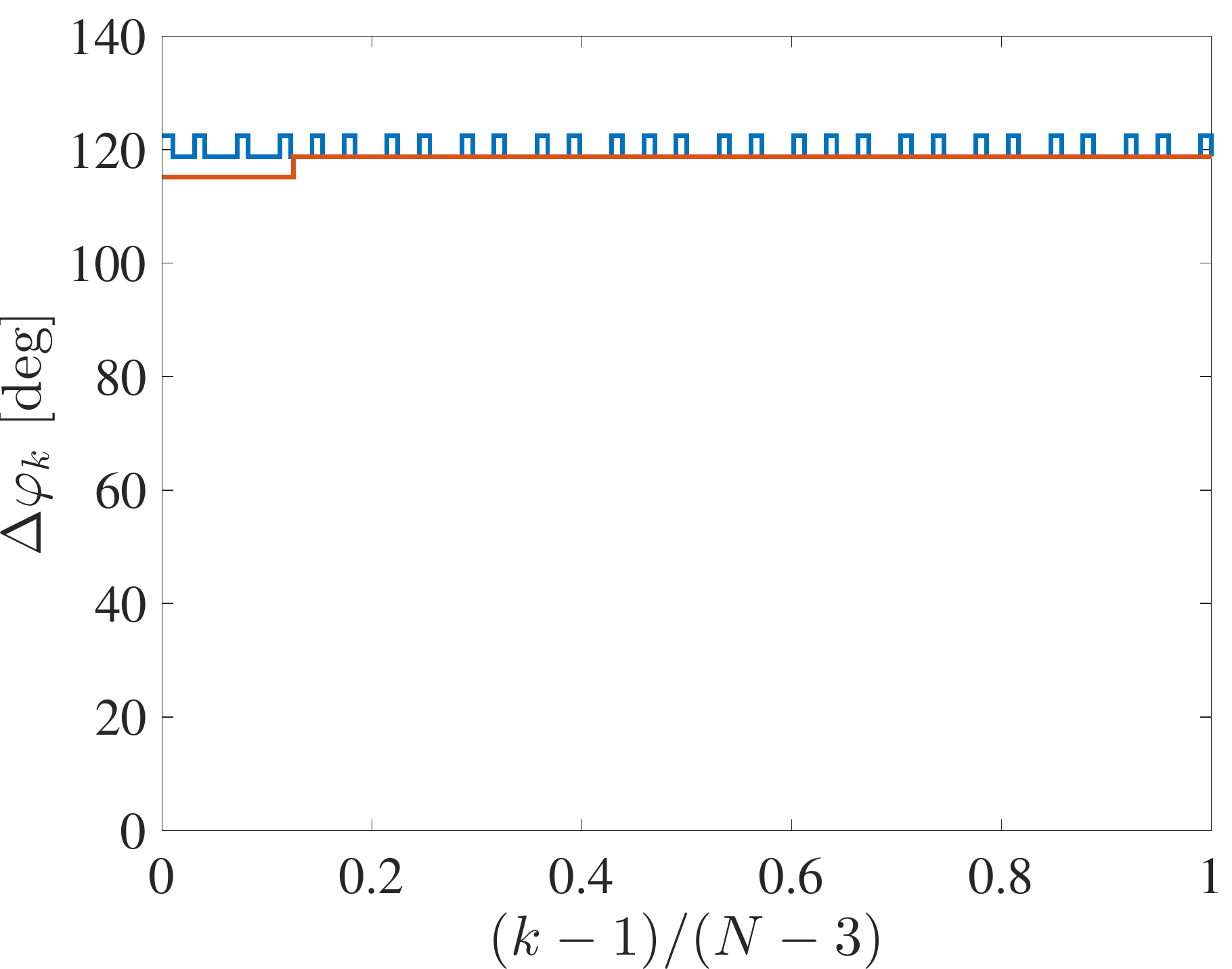}
\caption{}
\label{fig:phase101&11}
\end{subfigure}
 \caption{Optimal strokes for swimmers with $N=11$ (red) and $N=101$ (blue and blow-up) for $\Delta x=0.01 l$, where $l$ is the length of a single link. (a) A single snapshot of both swimmers during the optimal stroke. (b) Joint angle's amplitudes $a_k$ vs. (scaled) joint number $k$. (c) Phase differences $\Delta p_k$, defined by \eqref{eq:deltapk}, vs. (scaled) joint number $k$.}
 \label{Fig:11&101links}
 \end{figure}

\begin{figure}[h]
\centering
\begin{subfigure}{0.45\textwidth}
\includegraphics[width=\textwidth]{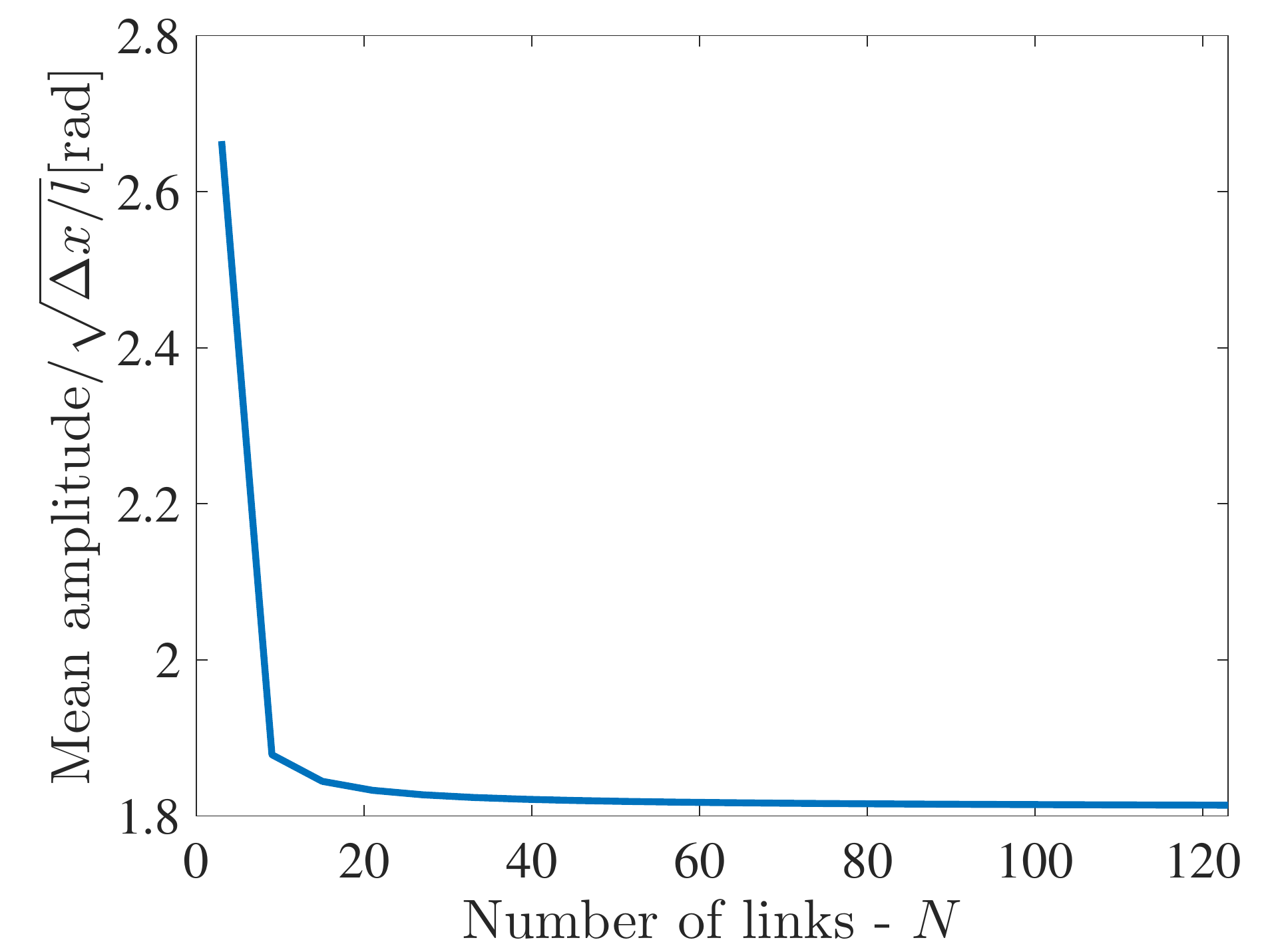}
\caption{}
\label{fig:mean_normalized_amp_vs_n}
\end{subfigure}
\begin{subfigure}{0.45\textwidth}
\includegraphics[width=\textwidth]{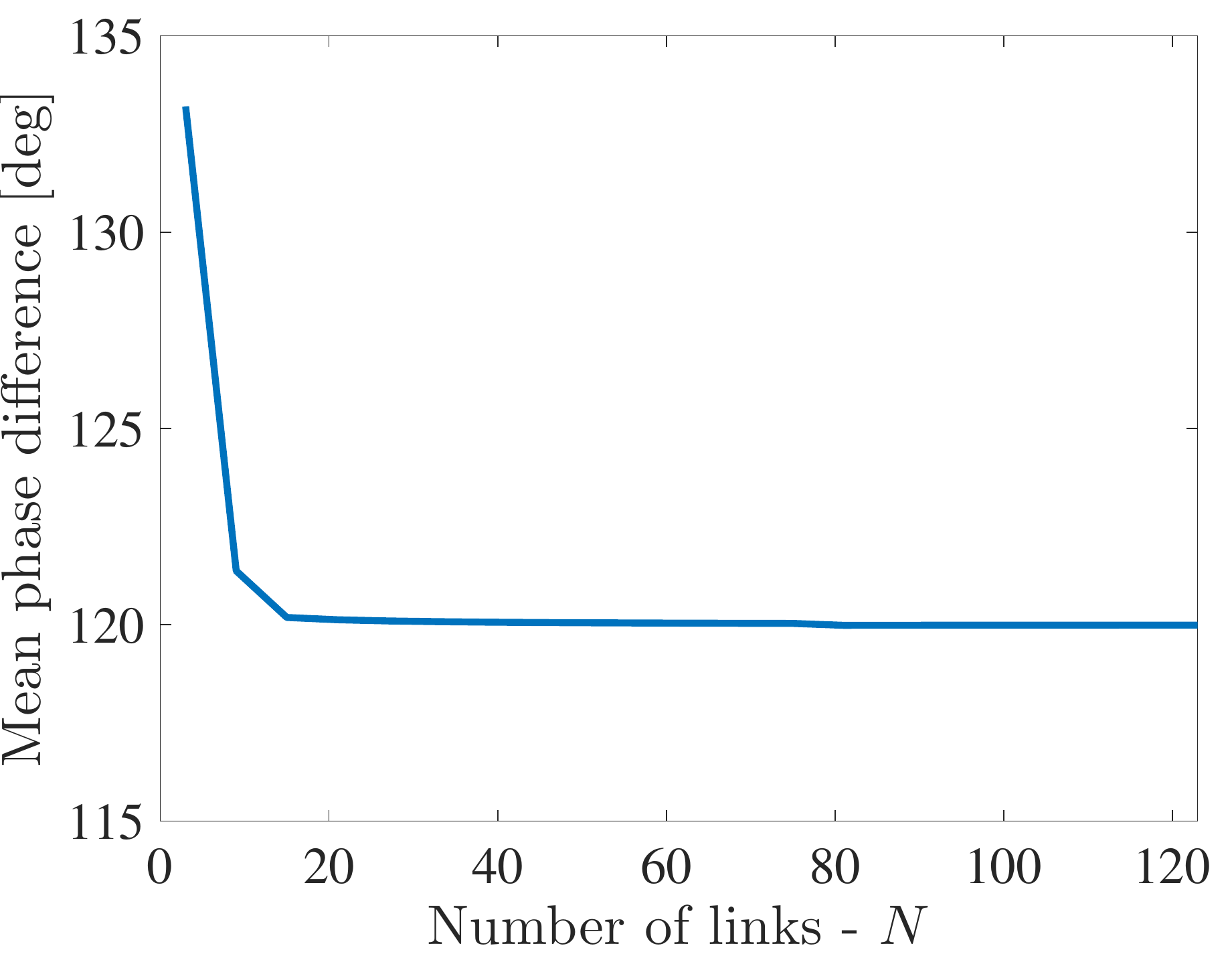}
\caption{}
\label{fig:meanphase_vs_n}
\end{subfigure}
 \caption{Optimal strokes for multi-link swimmers with large $N$ for $\Delta x=0.01 l$, where $l$ is the length of a single link.  (a)  Joint angle's amplitude $a_k$ averaged across all joints, as a function of $N$. (b) Phase difference $\Delta p_k$ averaged across all joints, as a function of $N$. }
 \label{Fig:function_of_N}
 \end{figure}

We now show results of optimal gaits for swimmers with $N=11$ and $N=101$, obtained by solving the eigenvalue problem  \eqref{eq:optimal_stroke}. Figure \ref{fig:snapshot_101&11} shows a single snapshot of both swimmers under the optimal gait for a prescribed displacement $\Delta x= 0.1l$. The figure indicates that the optimal gaits look like a travelling wave. In order to test this observation quantitatively, we rewrite the optimal gaits using the sinusoidal representation \eqref{eq:sin}, and compute the joint amplitudes $a_k$ and phase difference between consecutive joints
\begin{equation}
\Delta p_k=p_{k+1}-p_k \mbox{ mod }[360^\circ]\,, 
\label{eq:deltapk}
\end{equation}
where $p_k$ are given by \eqref{eq:sin}. The results of amplitudes and phase differences across the joints for both swimmers are plotted in Figures \ref{fig:amplitude101&11} and \ref{fig:phase101&11}, respectively. Remarkably, the amplitudes $a_k$ for both swimmers display a nearly identical and slightly non-uniform symmetric distribution along the swimmer's body, with a peak at the center (and up to $14\% $ decrease towards the ends), see Fig \ref{fig:amplitude101&11}. The phase difference between joints is very close to a uniform value of $120^\circ$, indicating a travelling wave with wavelength of three links. This is further confirmed when computing the optimal gait for increasing numbers of $N$ links for $\Delta x=0.01l$. Figures \ref{fig:mean_normalized_amp_vs_n} and \ref{fig:meanphase_vs_n} plot the gait's amplitude and phase difference, respectively, averaged across all swimmer's joints. Note that for a swimmer with discrete links, a waveform with length of integer number of $m$ links corresponds to a  phase difference of $\Delta \varphi=2 \pi /m$. Taking $m=1$ or $m=2$ results in time-reversible motion, hence the smallest possible integer is $m=3$. That is, the energy-optimal gait has the shortest possible wavelength. This result is in agreement with Taylor's observation in \cite{Taylor51}.

In hindsight, the emergence of traveling waves as optimal solutions for the control problem of a slender homogeneous swimmer whose shape is assumed to remain close to the rectilinear one, and swimming in a homogeneous fluid is not surprising. Consider for a moment the case in which the swimmer is of infinite length, but possesses a periodic shape (with period $Nl$). In particular, we look for strokes that are periodic in both space (with period $Nl$) and time (with period $T$). This system is translationally invariant with respect to shifts along the body axis. As a consequence, the initial problem \eqref{eq.dyn}, the optimal control problem \eqref{eq:problem}, the governing operators, and the resulting solutions possess the same symmetry. In particular, this entails that the matrices $\vecP_0$ and $\vecF^{\theta}_{\vecphi}-{\vecF^{\theta}_{\vecphi}}^T$ (and therefore $\vecM$) are circulant matrices, i.e. matrices $\vecX$ that are constant along their diagonals and whose entries $X_{ij}$ only depend on $i-j$ mod $[N-1]$. Such matrices share the same eigenvectors that are given by $\vecv^\pm_k=\frac{1}{\sqrt{N-1}}(1,\omega^\pm_k,(\omega^\pm_k)^2,\cdots,(\omega^\pm_k)^{N-2})^T$ with $\omega^\pm_k=e^{\pm 2ik\pi/{(N-1)}}$ for $k=1,\cdots,(N-1)/2$. Using \eqref{eq:phi_star_final} shows that the solution of the optimal control problem is therefore a traveling sinusoidal wave.
In the finite length case, the same is true modulo edge-effects, as it is apparent from Figure~\ref{Fig:11&101links}. Notice that a similar result has been obtained in the different but related context of crawling  of one-dimensional objects on solid surfaces or within a solid matrix \cite{Agostinelli}, where the optimality of actuation strategies based on peristaltic waves is discussed.

\begin{figure}[h]
\centering
\includegraphics[width=0.5\textwidth]{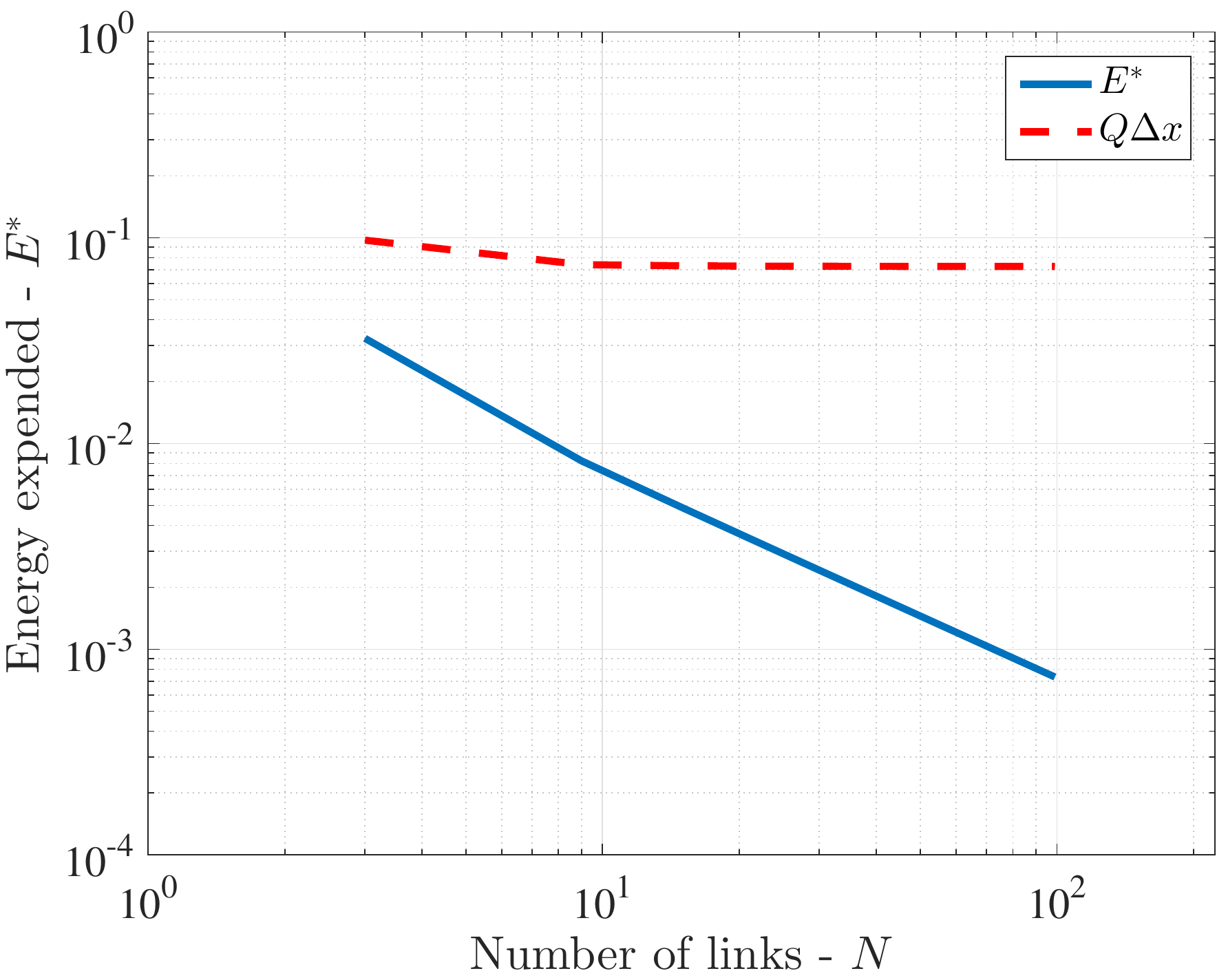}
 \caption{Log-log plots of the optimal energy $E^*$ (solid line) and scaled energy $Q$ (dashed line) for $\Delta x=0.01L$ as a function of $N$. Here $L=Nl$ is the total length of the swimmer and the energy is normalized by $c_t L^3 / T$.}
 \label{Fig:constant_L}
 \end{figure}

Finally, we fix the total length $L=Nl$ of a swimmer, compute the energy-optimal gait for moving a given displacement of $\Delta x = 0.01L$, and obtain the energy $E^*$ along this gait. Figure \ref{Fig:constant_L} shows a log-log plot of $E^*$ as a function of the links number $N$. Remarkably, it can be seen that for large $N$, the energy $E^*$ decays to zero as $1/N$. Analyzing this seemingly counter-intuitive behavior for large $N$ more closely (see Appendix) reveals that the optimal energy indeed scales as $E^* \sim \frac{\Delta x L^2}{N}$ for large $N$. This suggests that the a more suitable performance measure would be the scaled optimal energy defined as $Q=\frac{N E^*}{\Delta x L^2}$. This quantity, (multiplied by $\Delta x$ for better graphical visibility), is also plotted as a function of $N$ as the dashed line in Figure \ref{Fig:constant_L}, which indicates that it converges to a finite nonzero value at the limit of large $N$.

\section{Optimal strokes of large amplitude}\label{sec:large_amplitude}

In this section we remove the assumption of small angle amplitudes, and consider the problem of determining energetically-optimal strokes when the prescribed displacement in one cycle is no longer small. At the present state of our knowledge, this can only be done numerically.

We compute numerically the optimal gaits by utilizing the {\sc{Bocop}} toolbox of optimal control, which uses direct optimization methods by discretizing times and states \cite{bonnans2012Bocop}. {\sc{Bocop}} applies numerical integration of the full nonlinear dynamic equations \eqref{eq.dyn} and the energy formula \eqref{eq:energy_full}, without assuming small-amplitude strokes. We solve the problem for large prescribed displacements. In addition, we consider the behavior of the optimal solutions when the prescribed displacements become smaller and smaller, in order to obtain an independent check of our analysis of Section~\ref{sec:small_amplitude}, which is based on leading-order approximation when imposed displacements and angle amplitudes are small.
The discretized nonlinear optimization problem is solved by the {\sc{Ipopt}} solver \cite{wachter2006implementation} with {\sc{Mumps}} \cite{amestoy2001fully}, while the
derivatives are computed by sparse automatic differentiation with {\sc{CppAD}} \cite{bell2012cppad}.
In the numerical experiments, we used a midpoint (implicit 1st order) discretization with 100 time steps.

\subsection{Optimal gaits for Purcell's three-link swimmer}\label{sec:Purcell_large_amplitude}

\begin{figure}
\centering
\begin{subfigure}{0.35\textwidth}
\includegraphics[width=\textwidth]{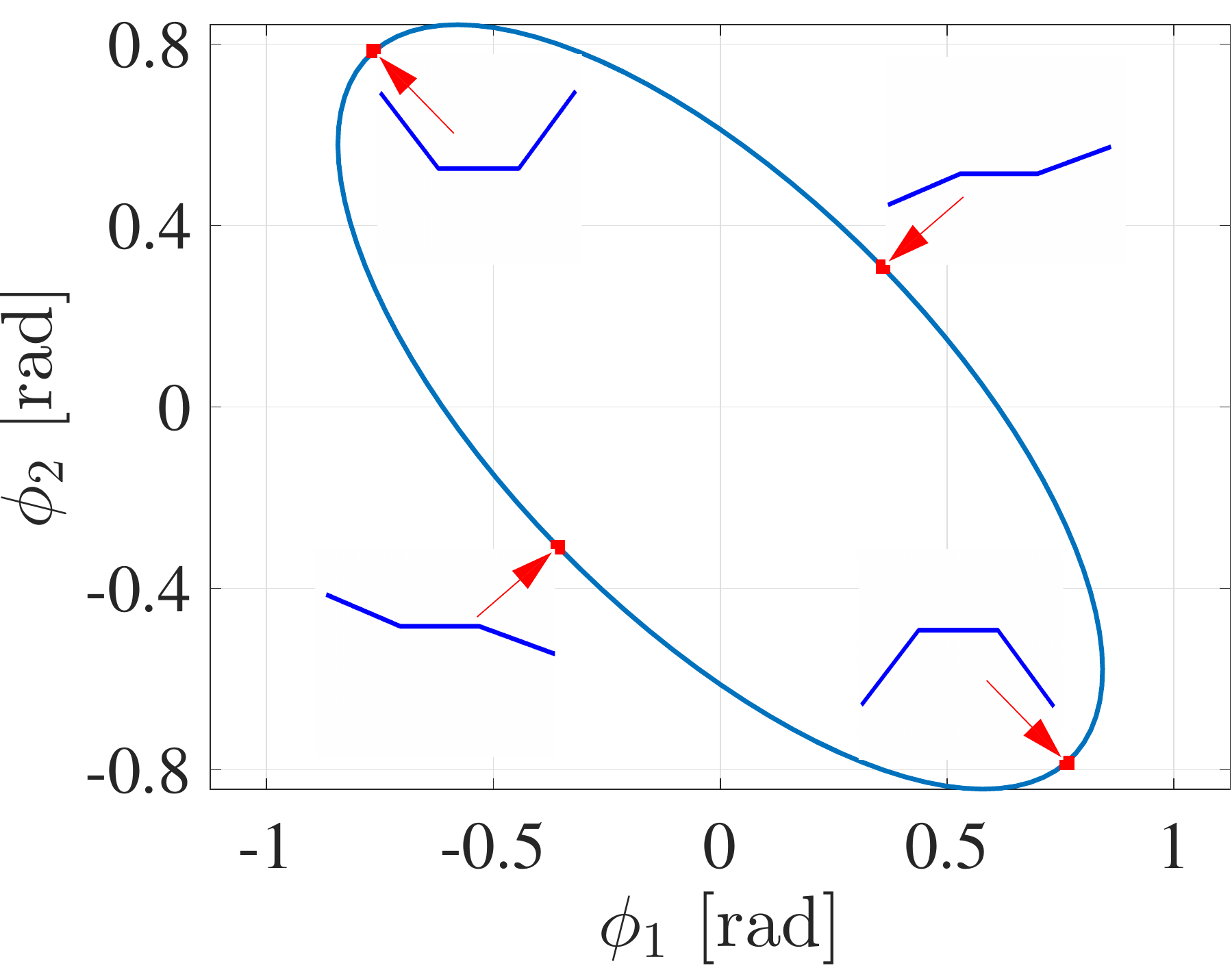}
\caption{}
\label{fig:optimal_gait_3link}
\end{subfigure}
\begin{subfigure}{0.4\textwidth}
\includegraphics[width=\textwidth]{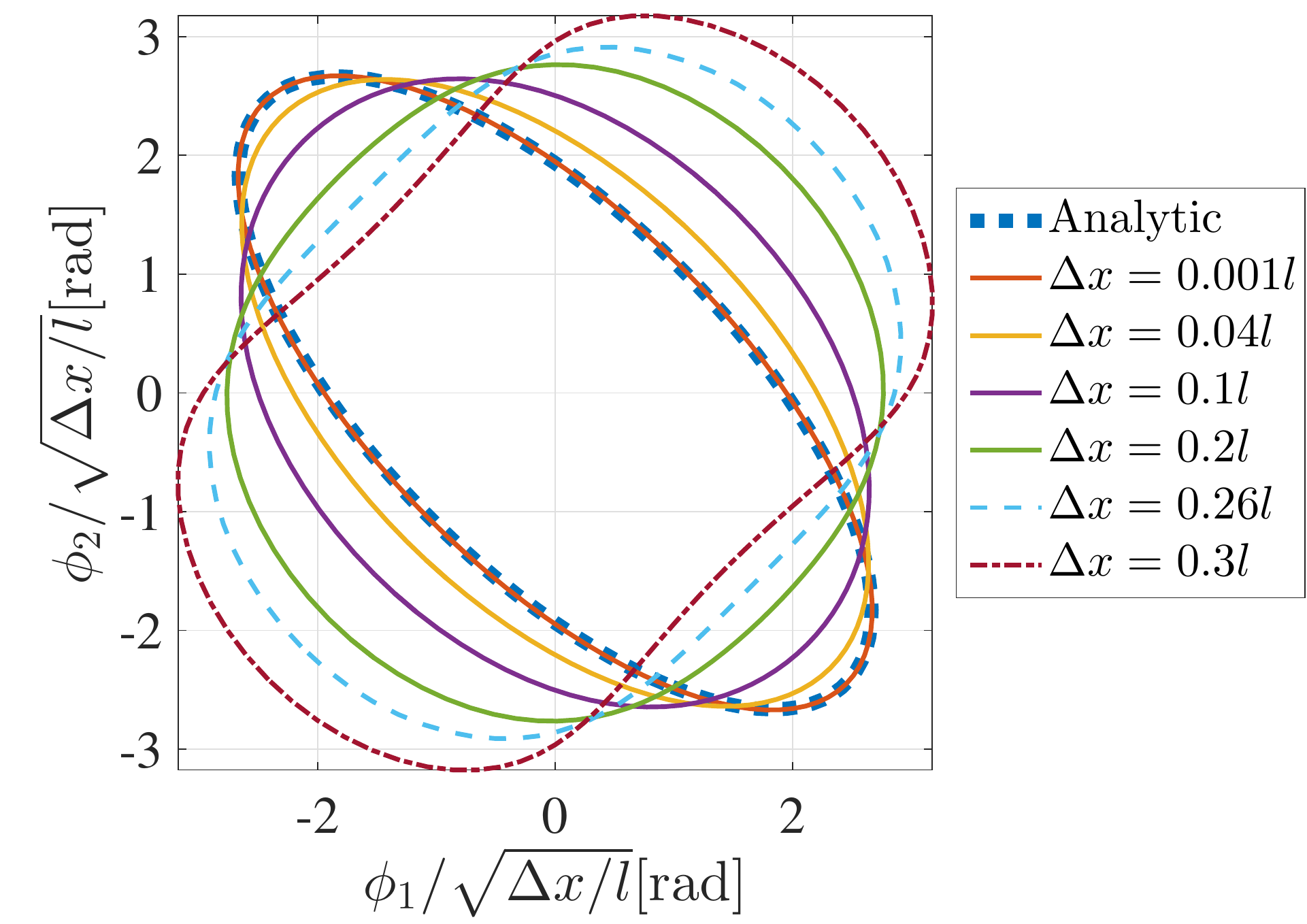}
\caption{}
\label{fig:Bocop_compare_3link}
\end{subfigure}
 \caption{Optimal strokes for Purcell's swimmer, plotted as loops in the plane of joint angles  $(\phi_1,\phi_2)$. (a) Optimal stroke obtained by our small-amplitude analysis, with snapshots of the swimmer's configurations. (b) Comparison with optimal strokes computed using {\sc{Bocop}} for different displacements $\Delta x$, angles $\phi_i$ are scaled by $\sqrt{\Delta x}$.}
 \label{Fig:3link_optimal_strokes}
 \end{figure}

The optimal gaits computed within the framework of the small angle amplitudes of Section~\ref{sec:Purcell} are shown in Fig. \ref{fig:optimal_gait_3link} as a trajectory in the plane of joint angles $(\phi_1,\phi_2)$ for $\Delta x = 0.1l$. Snapshots of the swimmer's configuration at quarter-period intervals are also illustrated on the plot.

For comparison, we computed energy-optimal gaits using {\sc{Bocop}} for different displacements $\Delta x$, and the resulting trajectories scaled by $\sqrt{\Delta x}$ are shown in Fig. \ref{fig:Bocop_compare_3link}. It can be seen from the plot that for small displacements $\Delta x$ the optimal gait obtained by {\sc{Bocop}} converges (after scaling) to the one obtained by the formula \eqref{eq:phi_star_final}. Note that we did not constrain the initial conditions of the swimmer in this computation with {\sc{Bocop}}, nor required any symmetry relations of the gait.
The only constraint is on zero net rotation and net displacement of magnitude $\Delta x$, without a specified direction. The agreement between the two methods of computation confirms the validity of our small-amplitude analysis. When the displacement $\Delta x$ is increased further, the energy-optimal gaits obtained with {\sc{Bocop}} begin to deviate significantly from the small-amplitude one. For $\Delta x = 0.26$, the optimal gait (dashed) coincides with the gait obtained by Tam and Hosoi  \cite{TamHosoi2007} that maximizes Lighthill's efficiency. This was calculated by considering the first two terms in a Fourier series expansion of the two joint angles. Notably, the long axis of this ellipse-like-shaped gait is ``skewed'' with respect to the long axis of the exact ellipse representing our small-amplitude energy-optimal gait.
The reason for this fundamental difference is the fact that optimization of Lighthill's efficiency as in \cite{TamHosoi2007} does not involve a constraint on the travelled distance. Therefore, it produces large displacements through large amplitude strokes, as discussed at the beginning of Section~\ref{sec:small_amplitude}.  

When $\Delta x$ is further increased to an upper limit of $\Delta x = 0.306$, the optimal gait (dash-dotted) deforms into the famous  peanut-shaped loop obtained in \cite{TamHosoi2007} as the maximal-displacement gait. Note that for large displacements, we had to add constraints on symmetries of the gait and initial conditions, otherwise {\sc{Bocop}} began to search for different gaits which are not ``simple loops''.

We now further analyze the motion of Purcell's swimmer using leading-order terms as studied in \cite{wiezel2016optimization}. The sinusoidal gait in \eqref{eq:sin} for the two joint angles can be rewritten as
\begin{equation} \label{eq.gait_phase}
\phi_1(t)=\vep \sin(t+\varphi/2) \, \;\; \phi_2(t)=\vep \sin(t-\varphi/2).
\end{equation}
That is, it has a stroke amplitude of $\vep$ and phase difference of $\varphi$. Using  resistive force theory and the leading order expansion in $\vep$, as explained in \cite{wiezel2016optimization}, the leading-order expressions for displacement $\Delta x$ and energy $E$ in a cycle under the gait \eqref{eq.gait_phase} are obtained as:
\begin{equation} \label{eq.leadXE}
\Delta x=\frac{5 \pi}{81}\vep^2 l \sin(\varphi) , \;\; E=\frac{2\pi c_t}{81}\vep^2 (11\cos(\varphi)+16) \; .
\end{equation}
Therefore, for sinusoidal gaits of the form \eqref{eq.gait_phase}, our constrained optimization of minimizing the energy to cover a given displacement reduces to minimizing $E / \Delta x$. This gives an optimization problem for a scalar function of $ \varphi$. Using elementary calculus, the minimum of this function is obtained at the optimal phase difference of $\varphi^*=\cos^{-1}(-11/16)=133.43^\circ$, and the resulting optimal gait \eqref{eq.gait_phase} is exactly identical to the one obtained by using the eigenvalue formulation in \eqref{eq:phi_star_final}, which is shown in Fig. \ref{fig:optimal_gait_3link}.

\subsection{Optimal gaits for 5-link swimmer}
\begin{figure}[htb]
\centering
\begin{subfigure}{0.24\textwidth}
\includegraphics[width=\textwidth]{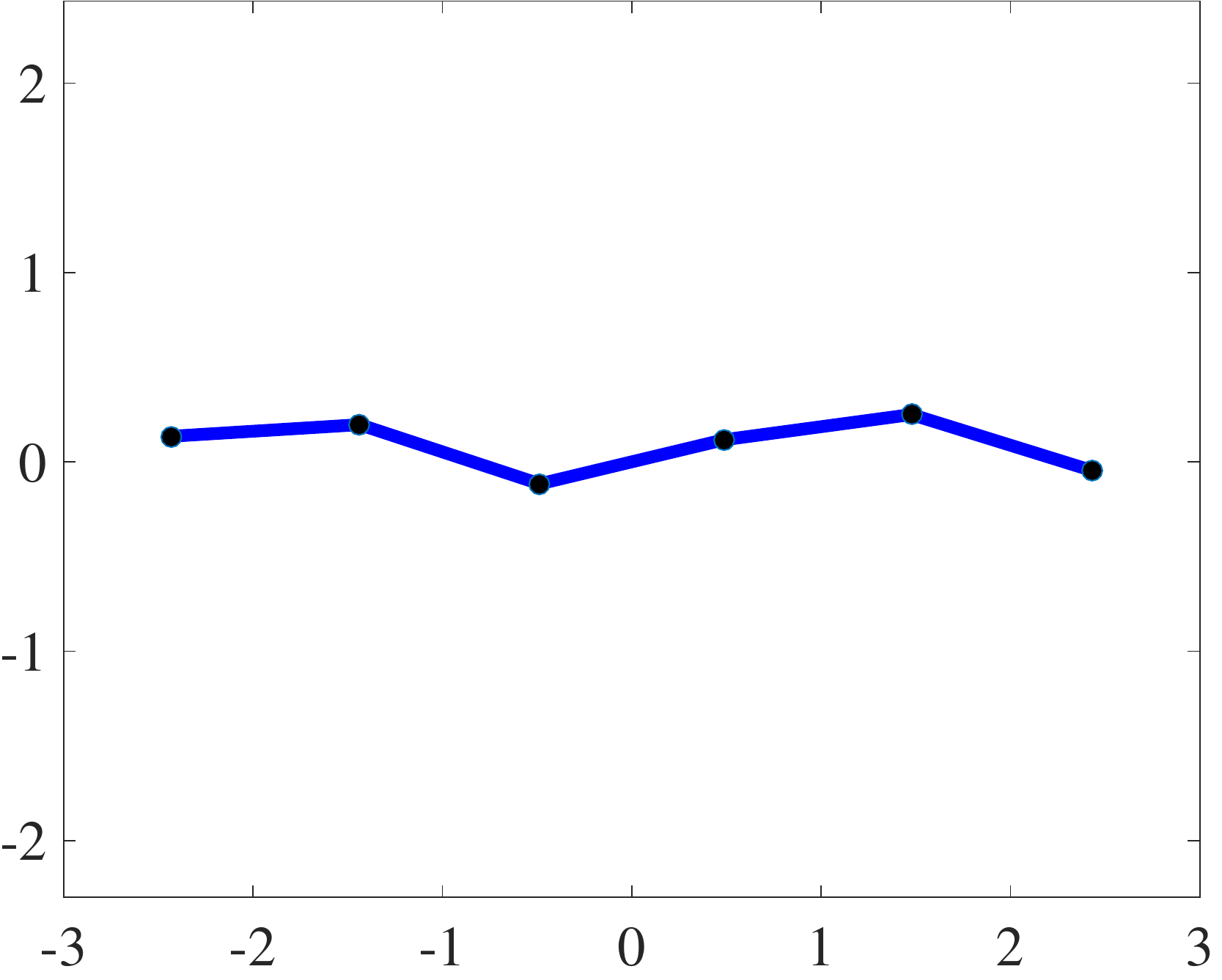}
\caption{}
\label{fig:5linksnap1}
\end{subfigure}
\begin{subfigure}{0.24\textwidth}
\includegraphics[width=\textwidth]{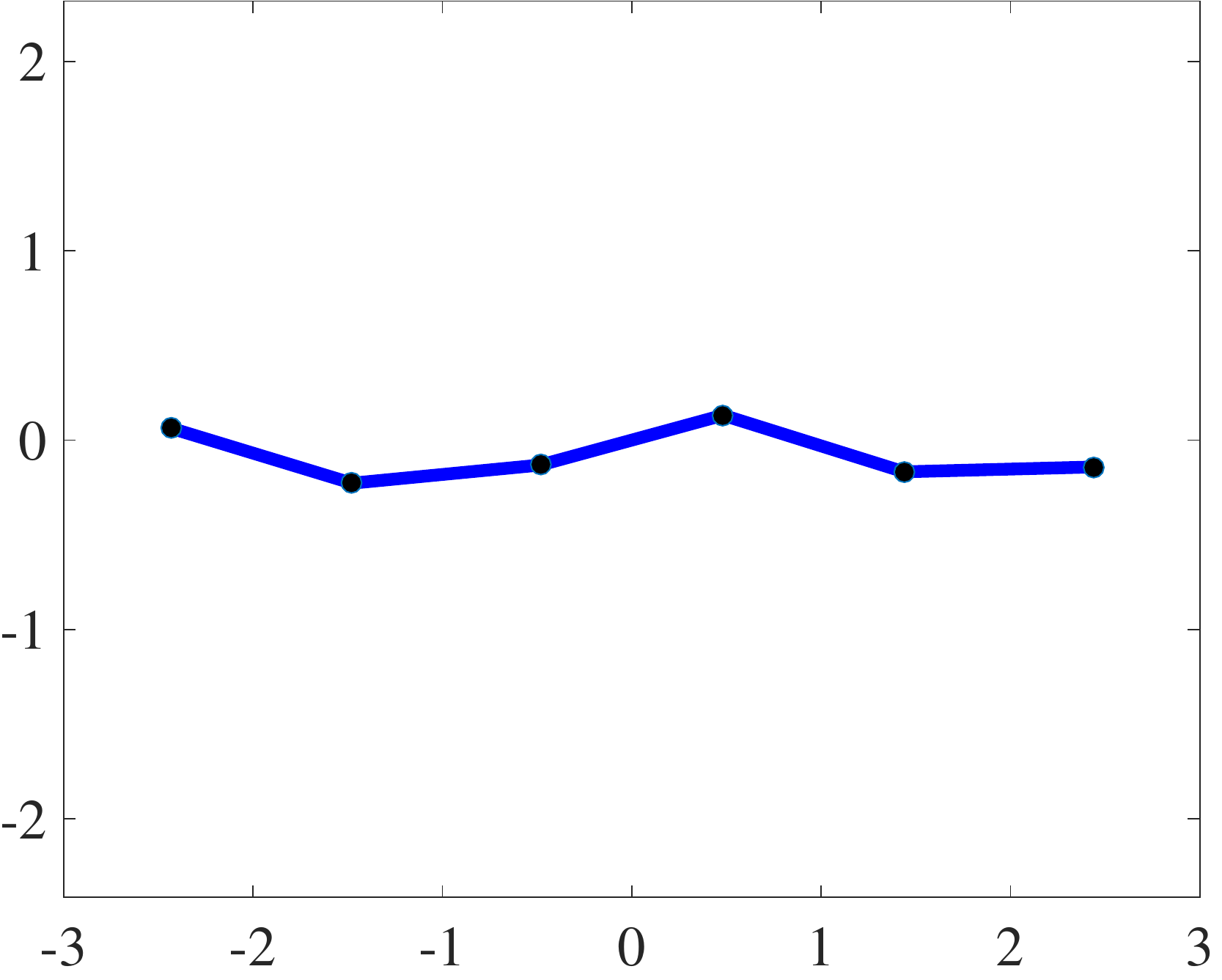}
\caption{}
\label{fig:5linksnap2}
\end{subfigure}
\begin{subfigure}{0.24\textwidth}
\includegraphics[width=\textwidth]{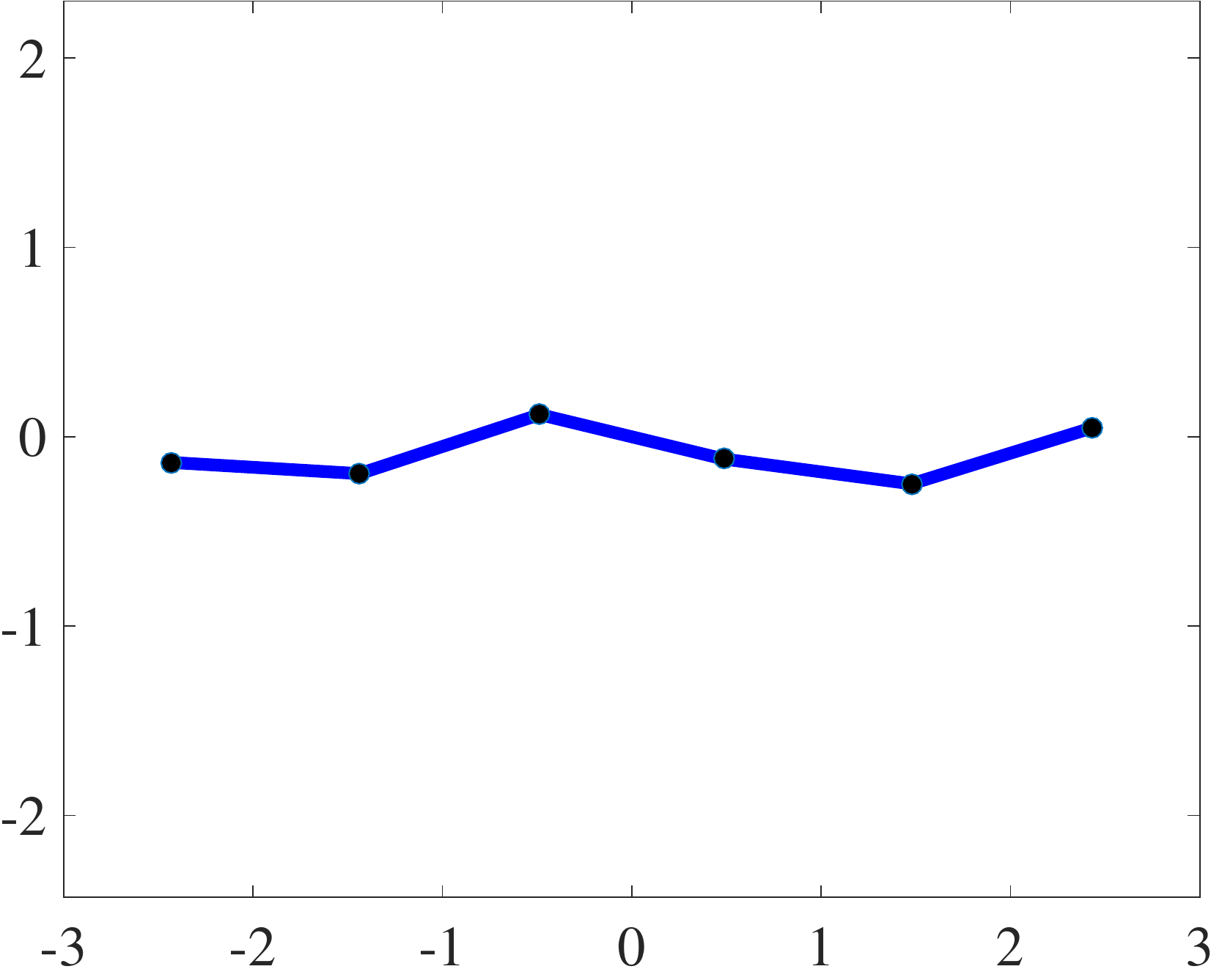}
\caption{}
\label{fig:5linksnap3}
\end{subfigure}
\begin{subfigure}{0.24\textwidth}
\includegraphics[width=\textwidth]{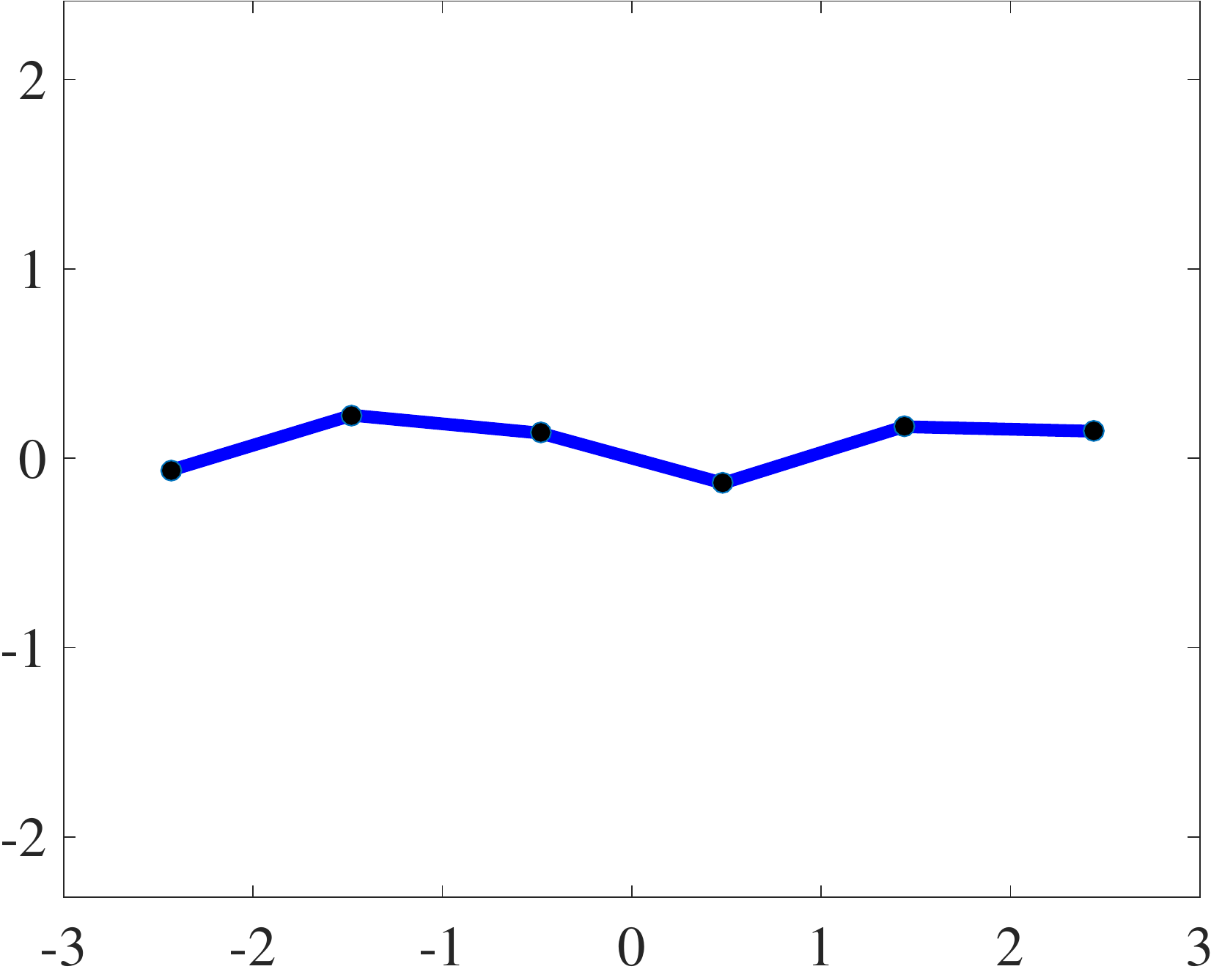}
\caption{}
\label{fig:5linksnap4}
\end{subfigure}
\begin{subfigure}{0.24\textwidth}
\includegraphics[width=\textwidth]{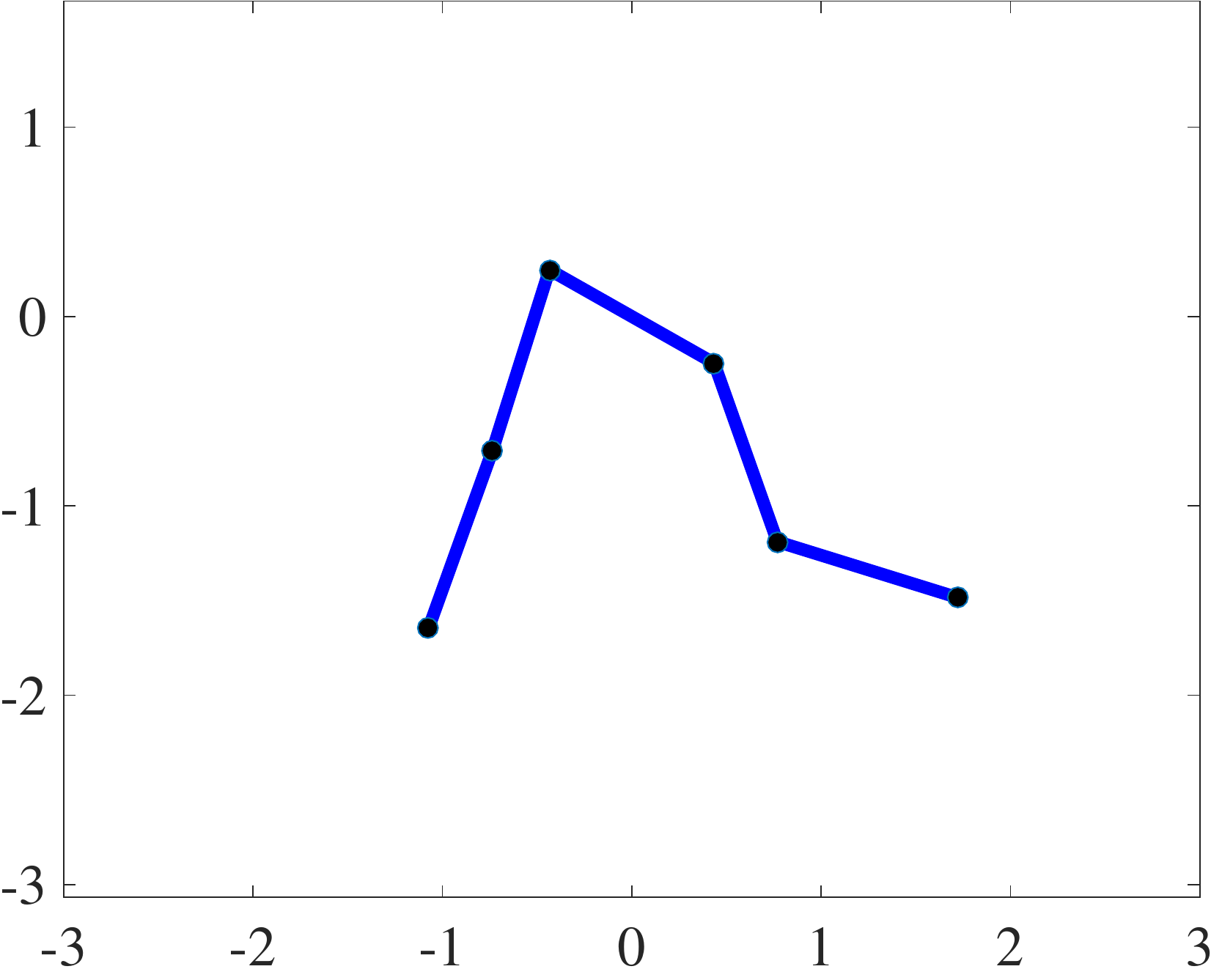}
\caption{}
\label{fig:5linksnap_large_amp1}
\end{subfigure}
\begin{subfigure}{0.24\textwidth}
\includegraphics[width=\textwidth]{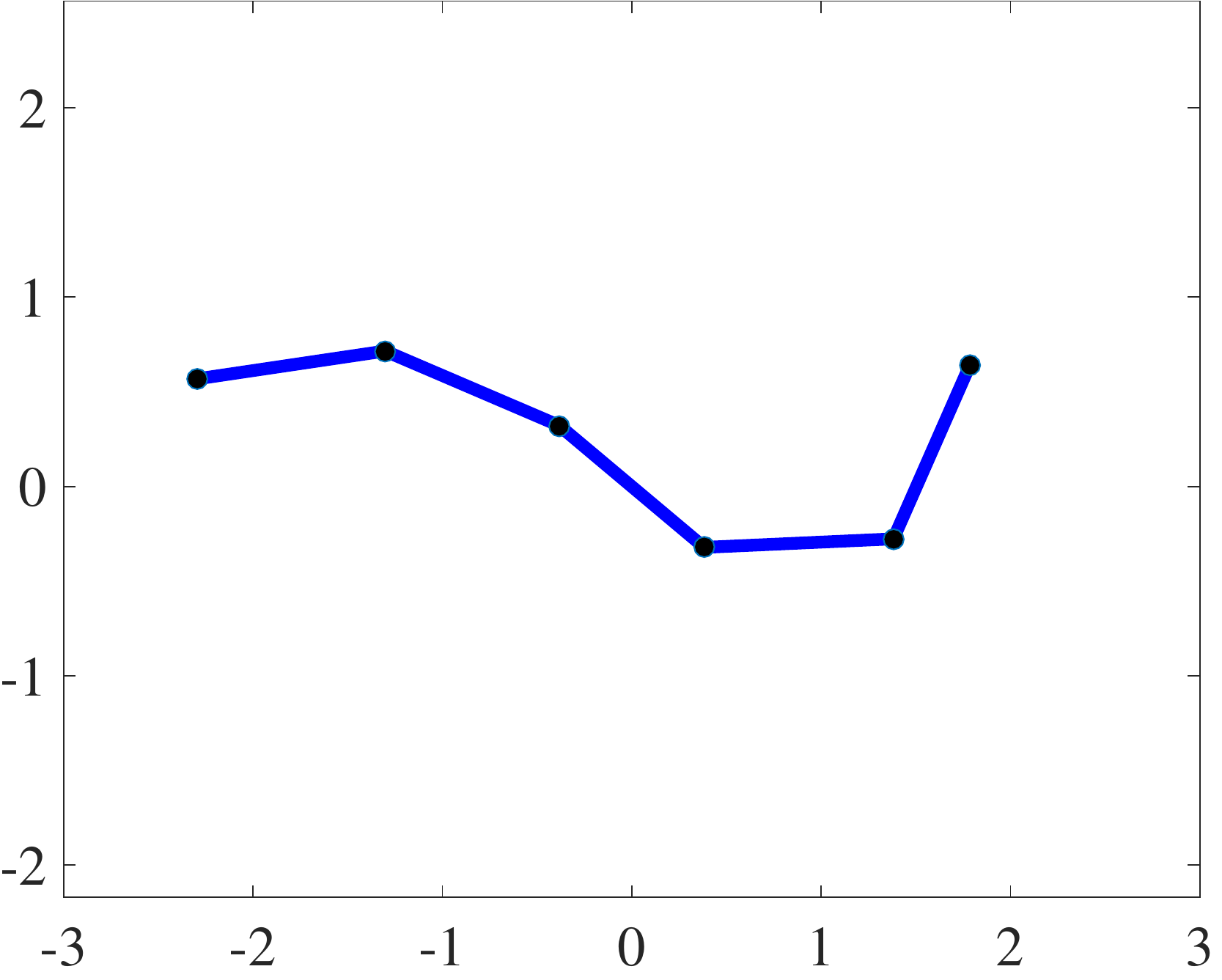}
\caption{}
\label{fig:5linksnap_large_amp2}
\end{subfigure}
\begin{subfigure}{0.24\textwidth}
\includegraphics[width=\textwidth]{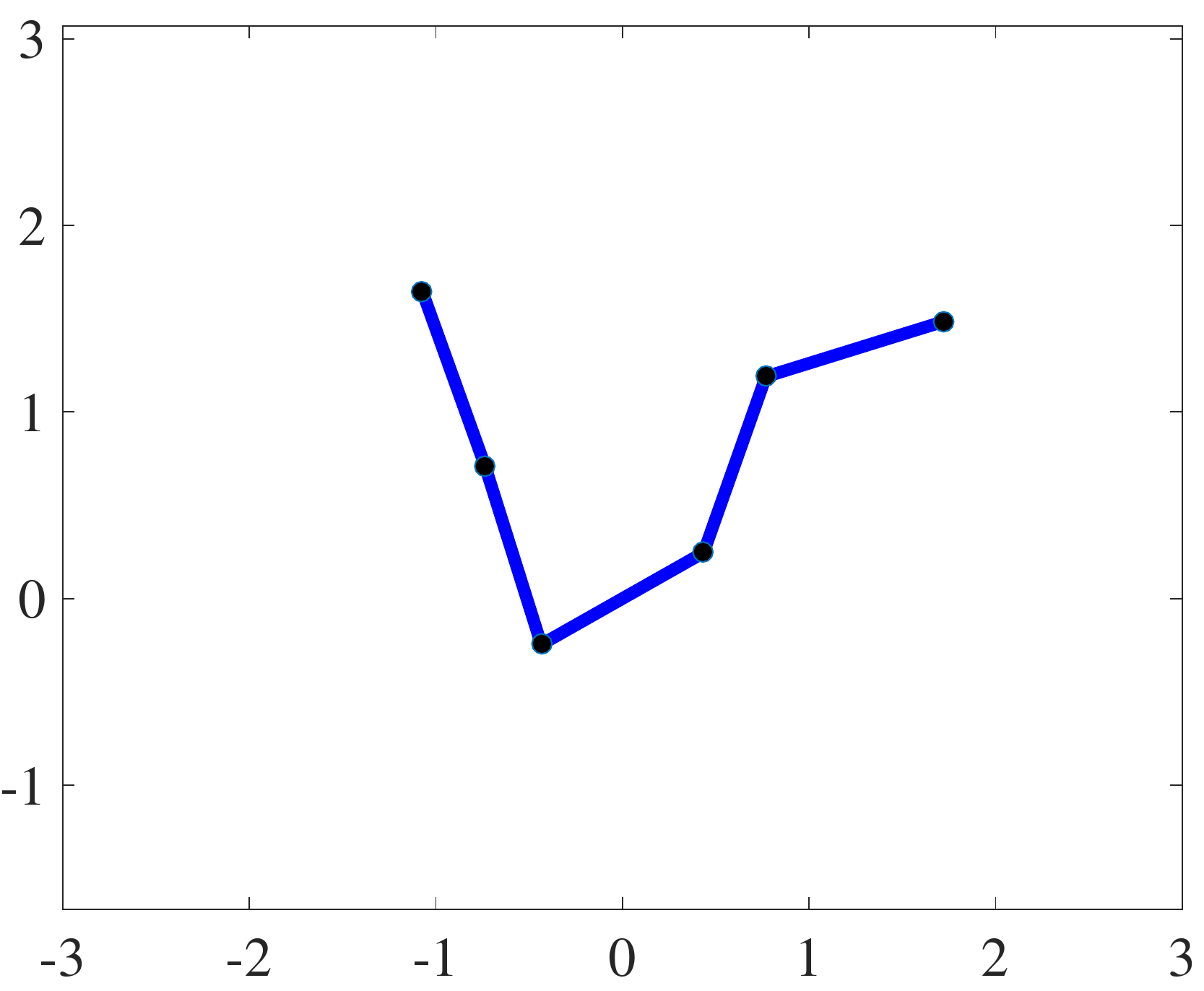}
\caption{}
\label{fig:5linksnap_large_amp3}
\end{subfigure}
\begin{subfigure}{0.24\textwidth}
\includegraphics[width=\textwidth]{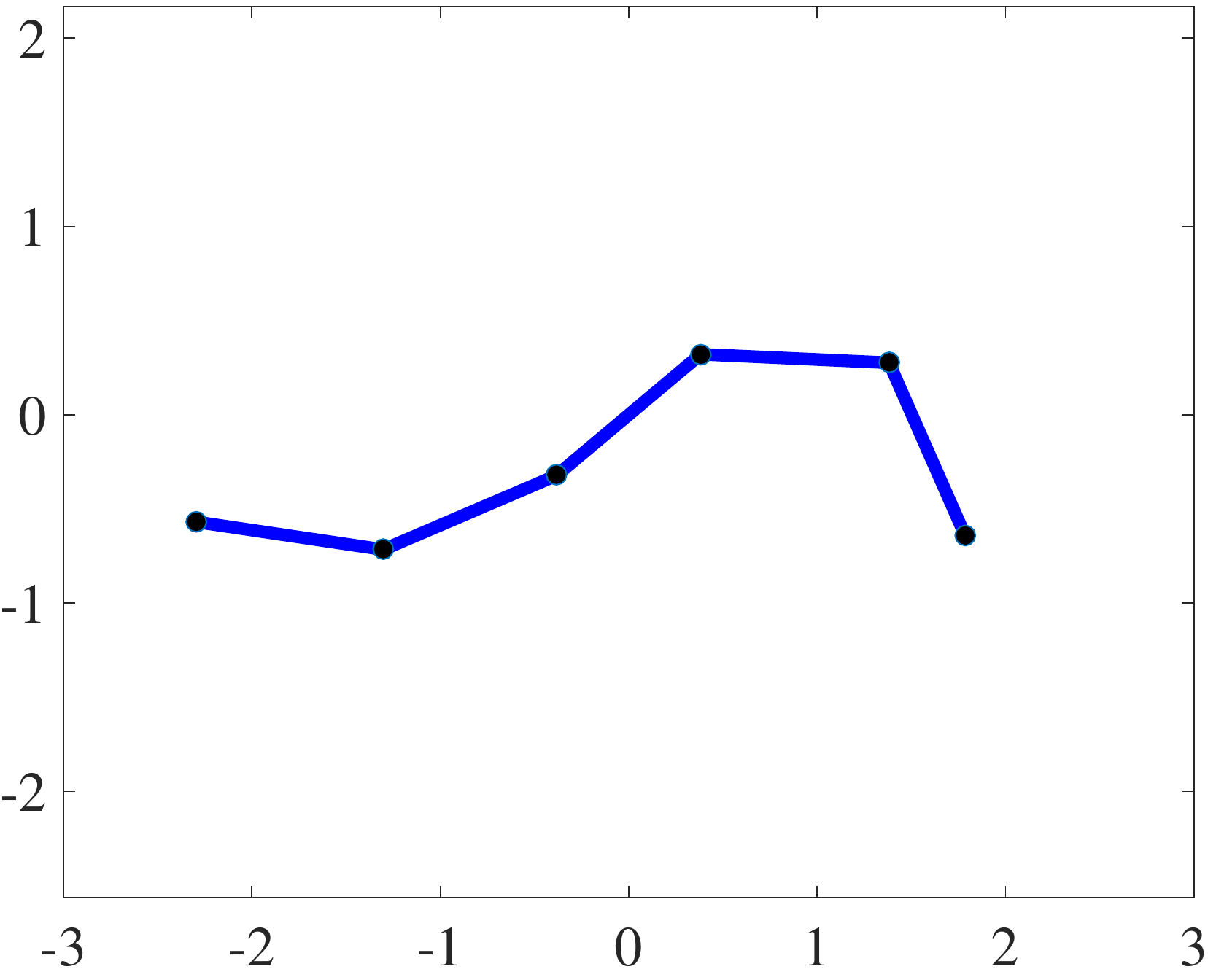}
\caption{}
\label{fig:5linksnap_large_amp4}
\end{subfigure}
 \caption{Snapshots of five-link swimmer during the optimal stroke. (a)-(d): analytical solutions for $\Delta x=0.1l$; (e)-(h): numerical solutions for $\Delta x=l$.}
 \label{Fig:5linksnap}
 \end{figure}

\begin{figure}
\centering
\begin{subfigure}{0.45\textwidth}
\includegraphics[width=\textwidth]{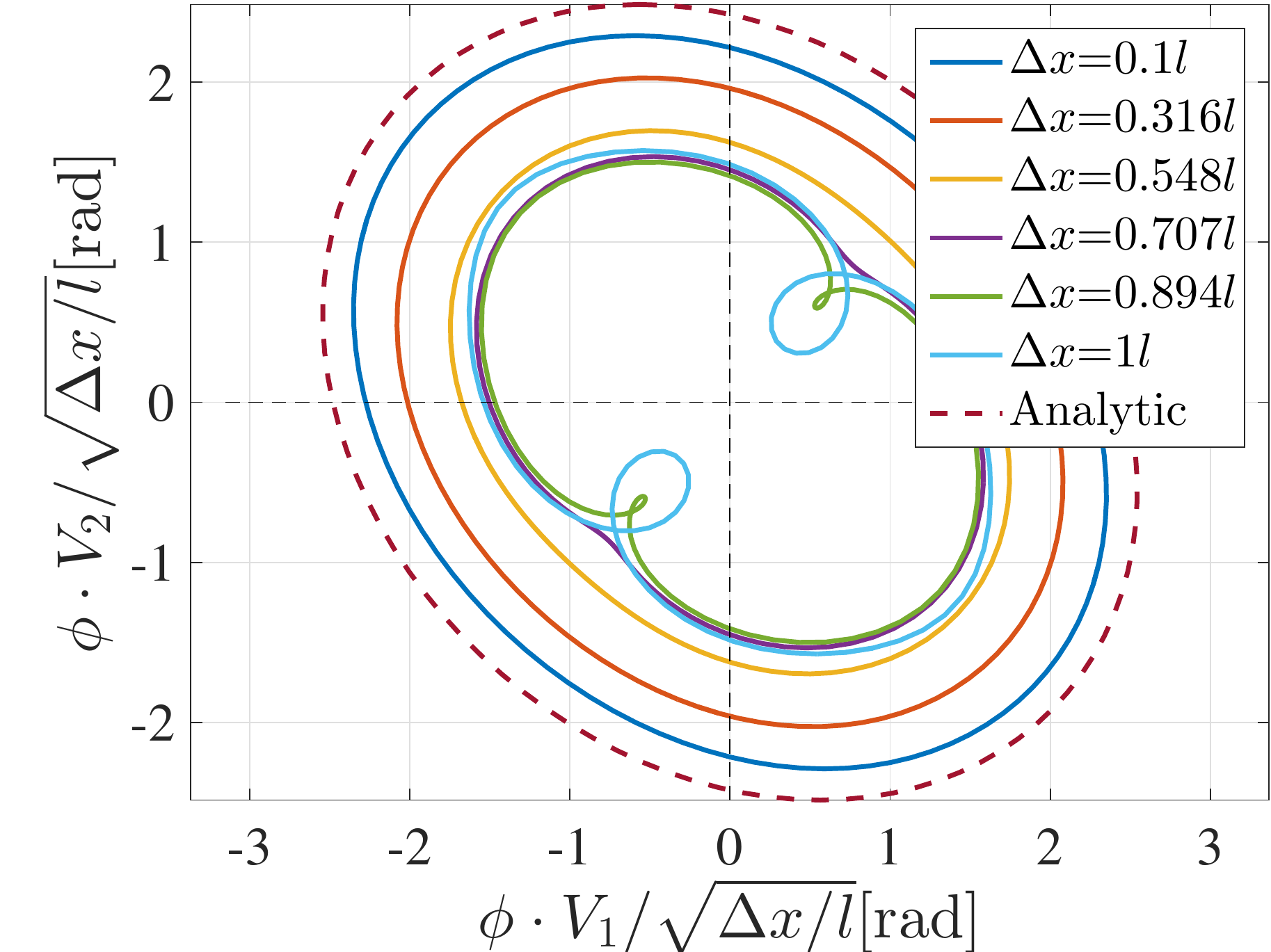}
\caption{}
\label{fig:Bocop_compare_5link}
\end{subfigure}
\begin{subfigure}{0.45\textwidth}
\includegraphics[width=\textwidth]{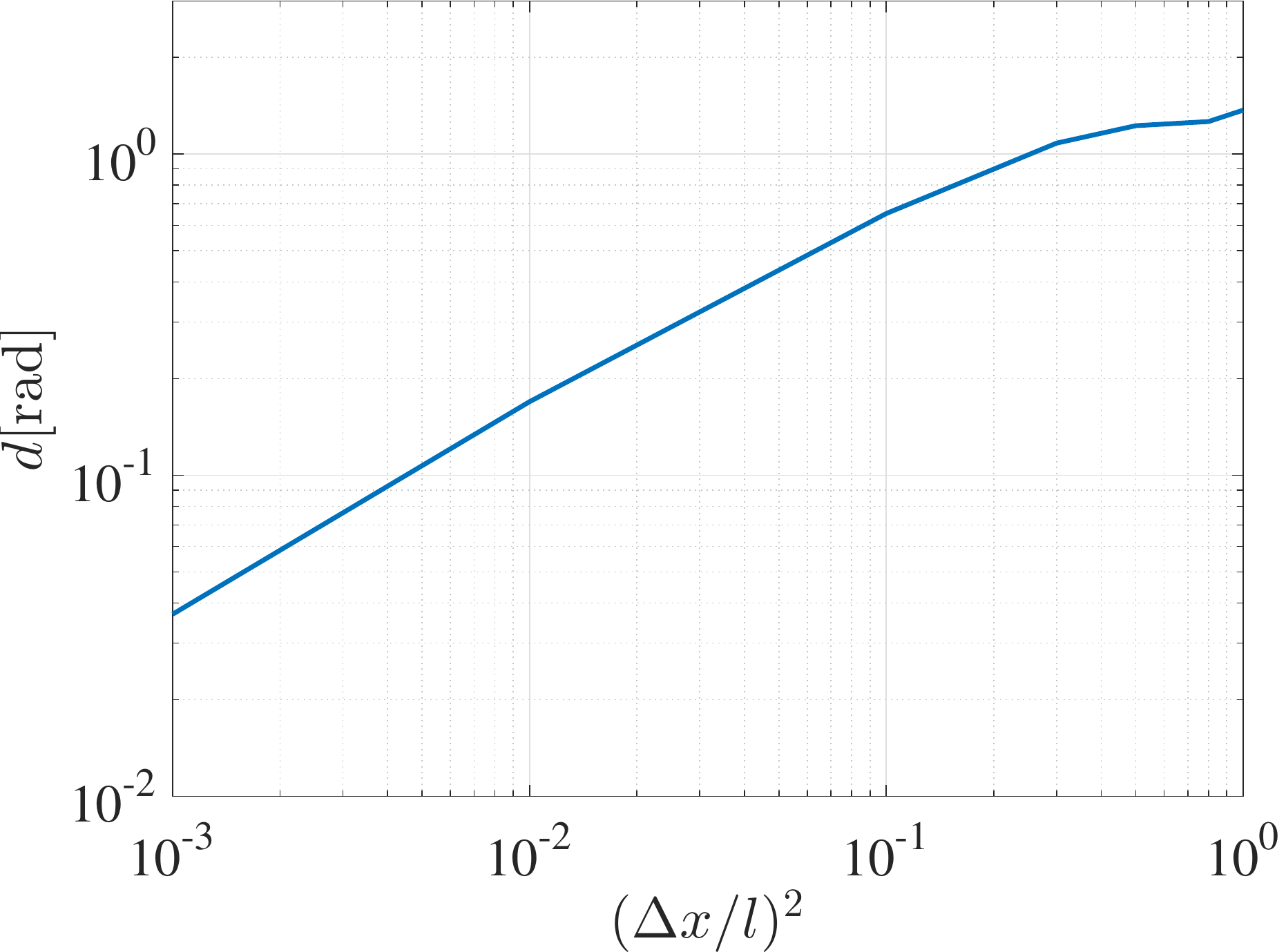}
\caption{}
\label{fig:distance_from_plane}
\end{subfigure}
 \caption{Comparison of the small-amplitude optimal stroke of the five-link swimmer with {\sc{Bocop}} computations for different displacements $\Delta x$. (a) Projection of the strokes onto the plane $S$, angles $\phi_i$ are scaled by $\sqrt{\Delta x}$ ($V1,V2 \in \mathbb{R}^4$ are two orthonormal basis vectors spanning the plane  $S$). (b) Log-log plot of the maximal distance $d$ of optimal strokes obtained with {\sc{Bocop}} from the plane $S$, as a function of  displacement $\Delta x$.}
 \label{Fig:5link_optimal_strokes}
 \end{figure}

We now show a comparison between our analytical formulation of optimal gaits in \eqref{eq:phi_star_final} and numerical optimization using {\sc{Bocop}} for the five-link swimmer, whose space of joint angles is four-dimensional. Snapshots  at quarter-period times of the swimmer's configuration  for the analytical optimal gait corresponding to $\Delta x=0.1l$  
are shown in Figure \ref{Fig:5linksnap} and compared to the numerical solutions obtained for $\Delta x=l$. According to our small-amplitude analysis, the optimal gait is planar  (it lies in the two-dimensional linear subspace $S$ spanned by eigenvectors associated with the pair of imaginary eigenvalues of $\vecM$ corresponding to the maximal magnitude $\mu_{\scriptscriptstyle M}$.)  In order to compare further our analytical optimal gaits with the results of {\sc{Bocop}} computations, we first plot in Figure \ref{fig:Bocop_compare_5link} the projections onto the planar subspace $S$  of the four-dimensional optimal gaits. These optimal gaits were numerically computed by {\sc{Bocop}} for different values of $\Delta x$ and then scaled by $\sqrt{\Delta x}$ for comparison with the analytical optimal gaits. In order to test the theoretical prediction that optimal gaits should lie within the two-dimensional subspace $S$, we computed the maximal Euclidean distance $d$ of each optimal gait in $\mathbb{R}^4$ from the plane $S$. Figure \ref{fig:distance_from_plane} shows a log-log plot of this distance $d$ as a function of $\Delta x$. It can be seen that for small displacements this distance decays to zero as $(\Delta x)^{3/2}$, indicating that the optimal gaits obtained numerically using {\sc{Bocop}} are indeed converging to planar loops lying within $S$, in agreement with the prediction of our asymptotic analysis. 

\begin{figure}[h]
\centering
\begin{subfigure}{0.31\textwidth}
\includegraphics[width=1\textwidth]{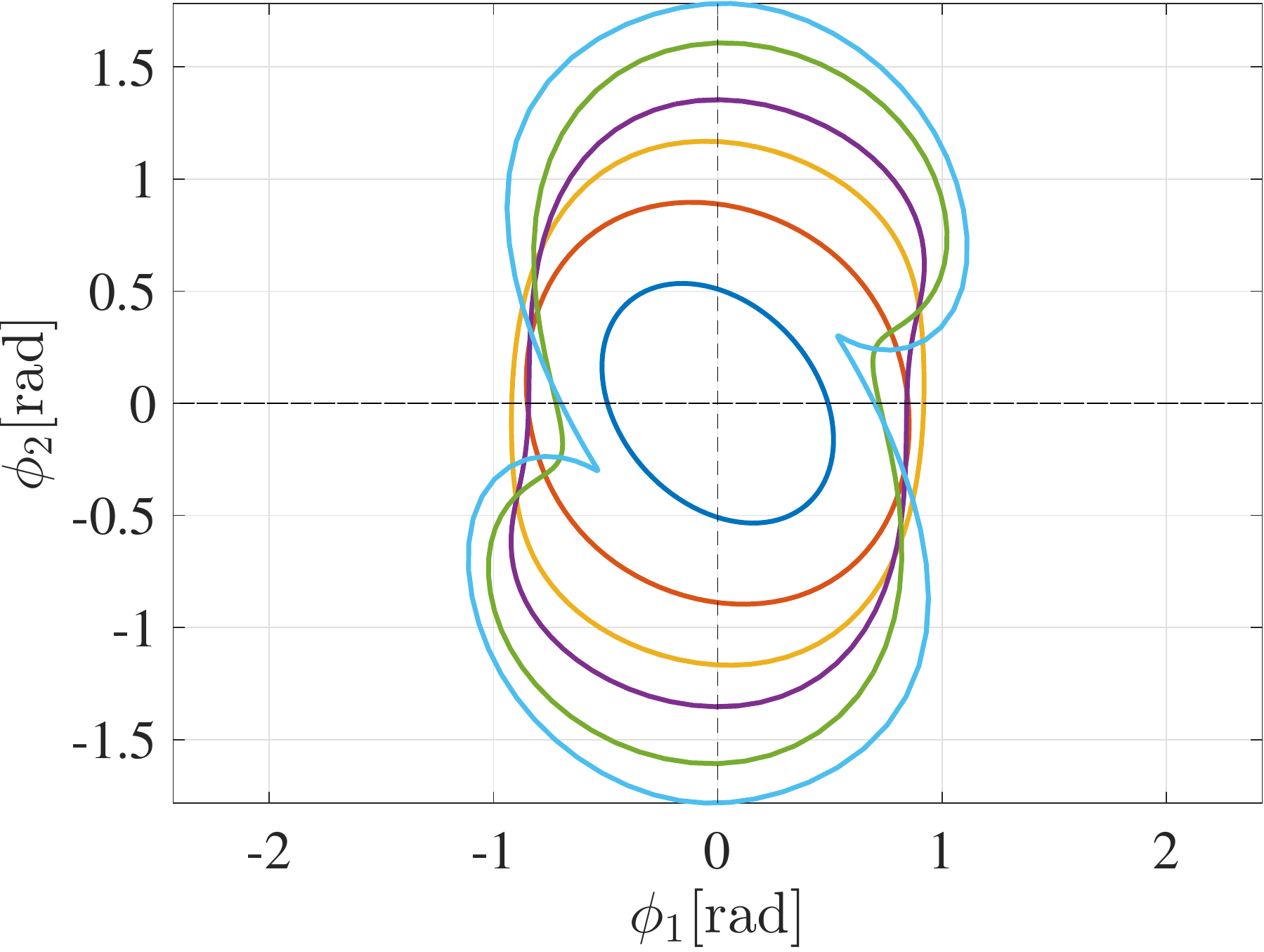}
\caption{}
\label{fig:phi1phi2}
\end{subfigure}
\begin{subfigure}{0.31\textwidth}
\includegraphics[width=\textwidth]{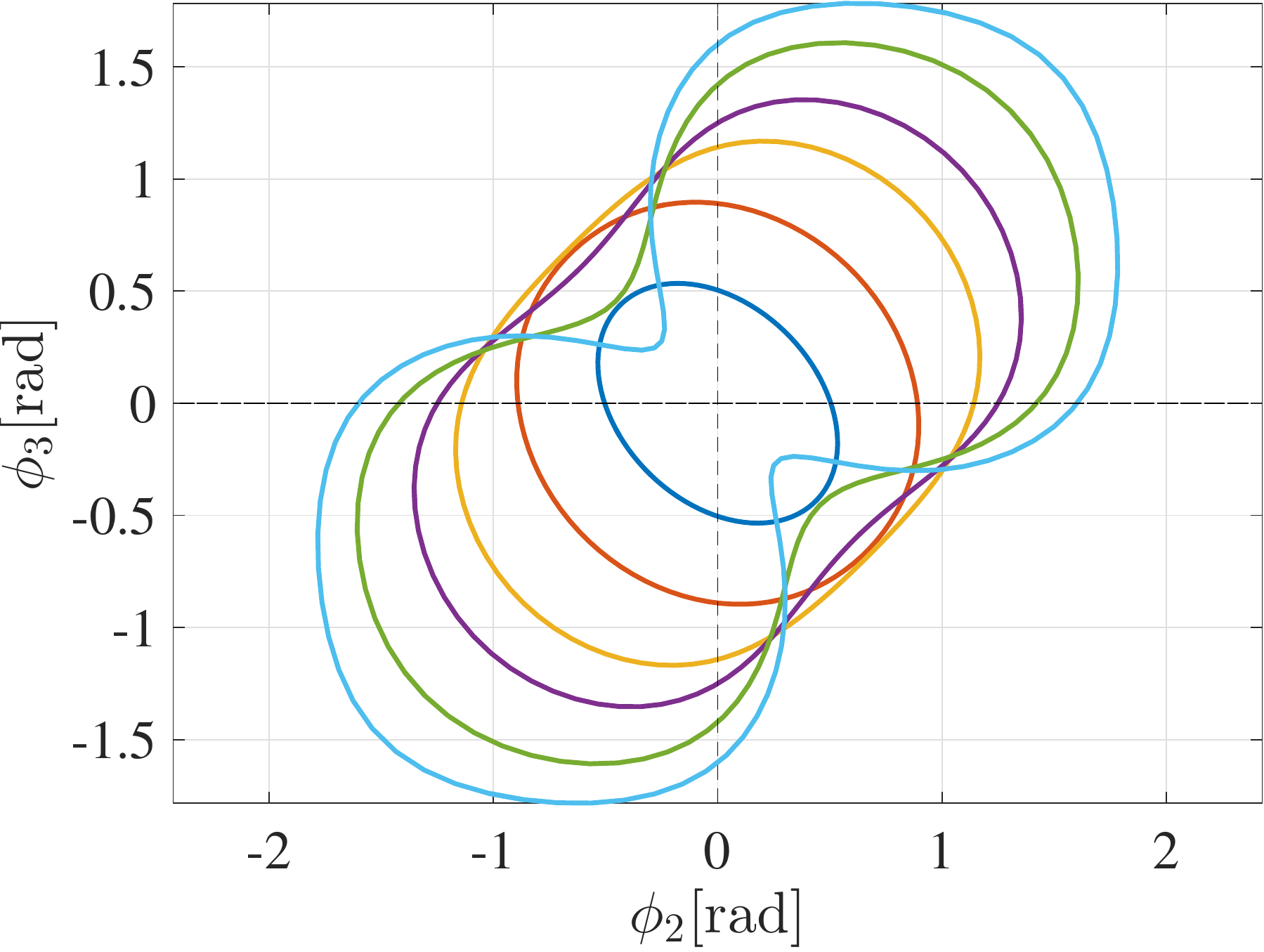}
\caption{}
\label{fig:phi2phi3}
\end{subfigure}
\begin{subfigure}{0.31\textwidth}
\includegraphics[width=\textwidth]{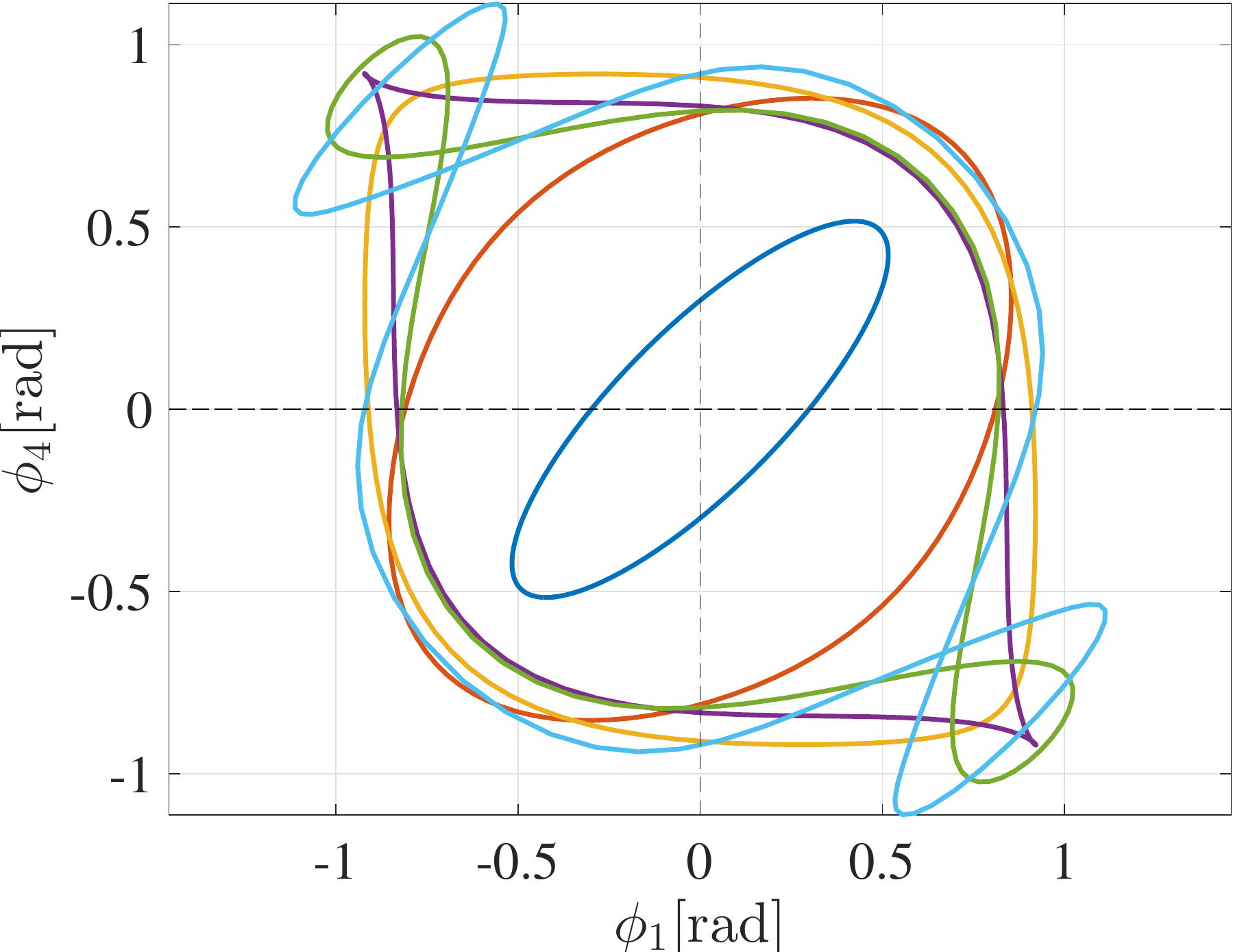}
\caption{}
\label{fig:phi1phi4}
\end{subfigure}
 \caption{Optimal strokes of the five-link swimmer obtained with {\sc{Bocop}} computations for different displacements $\Delta x$. (a) $\phi_1$ versus $\phi_2$. (b) $\phi_2$ versus $\phi_3$. (c) $\phi_1$ versus $\phi_4$.}
 \label{Fig:5link_optimal_strokes_projected}
 \end{figure}

Moving to the discussion of the optimal gaits of large amplitude, the fact that we are considering a system whose shape space is four-dimensional enables us to address and explore the deviations from planarity of the optimal loops.
The computed optimal strokes are shown in the bottom panel of Fig.~\ref{Fig:5linksnap} 
and in Fig.~\ref{Fig:5link_optimal_strokes_projected},  where we plot several different projections, by plotting the joint evolution of several pairs of shape variables. While, in the regime of small displacements, we do recover the elliptical loops predicted by our theory, the shape of the strokes for large displacements shows large discrepancies, and the geometry of the strokes is difficult to understand because of  their high-dimensionality. One notable feature of Fig.~\ref{Fig:5linksnap}  is that, in the large angle amplitude regime, optimal strokes consist of undulating shapes (wave-forms) such that the distance between successive peaks is of the order of the whole swimmer length, see Fig.~\ref{fig:5linksnap_large_amp4}, as it is in fact observed in biological systems such as sperm cells \cite{Gaffney2011,Gaffney2017}.   

\begin{figure}[h]
\centering
\includegraphics[width=\textwidth]{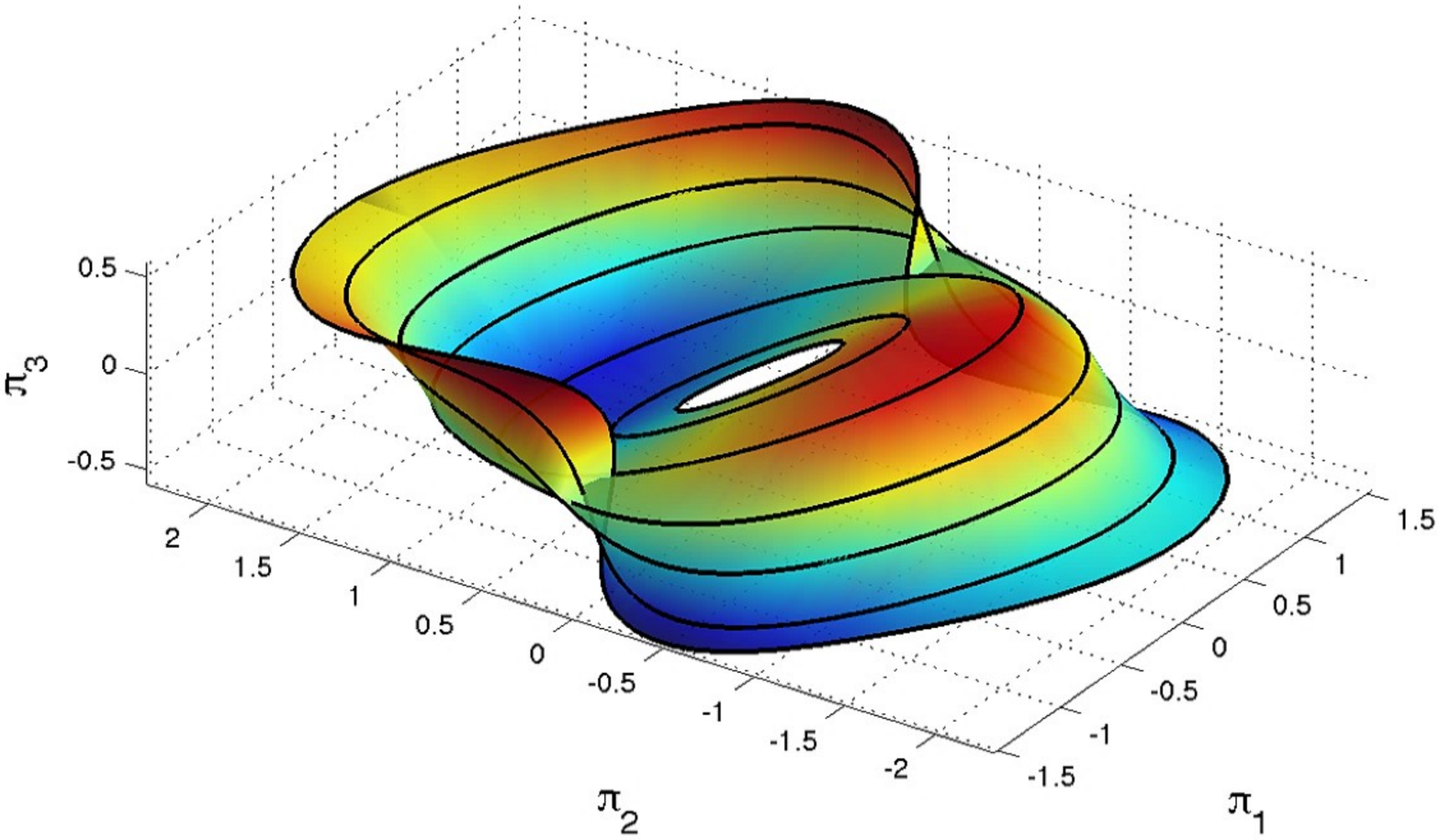}
\caption{Three-dimensional representation of the optimal loops for the 5-link-swimmer as a function of the amplitude of the prescribed displacement $\Delta x$, in normalized coordinates $(\pi_1,\pi_2,\pi_3)$. The 3D picture is produced using the {\sc{Isomap}} approach and the optimal strokes for different prescribed displacements are plotted in black. It is therefore a three dimensional non-linear projection of the origianl angles in 4D. As the displacement goes to 0, the strokes converge to a planar ellipse, as the one obtained in the small displacement limit. In turn, when the prescribed displacement increases, the strokes are non-planar and their geometry is complex.}
\label{fig:isomap}
\end{figure}

In order to give a concrete and visual representation of the emergence of non-planarity when angles and displacements become large, we derive from the computed four-dimensional loops a dimensionally-reduced representation in three dimensions, using the {\sc{Isomap}} approach \cite{isomap,ArroyoDeSimone12}.
Figure~\ref{fig:isomap} shows clearly that the loops are non-planar, and that interesting features of their two-dimensional projections such as cusps and crossings are really an outcome of the projection of higher dimensional, non self-intersecting curves. This interesting behavior is entirely new with respect to the results currently available in the literature for the $N=3$ case of Purcell's swimmer whose shape space is two-dimensional, and whose optimal strokes are necessarily planar curves. In addition, we see clearly that, as the imposed displacement becomes small, the loops converge to the planar ellipses of the small amplitude approximation of Section~\ref{sec:small_amplitude}.

These results open the new and unexplored question of characterizing the geometry of optimal loops for swimmers with shape space of large dimensions, when the amplitude of the joint angles is allowed to become large. Conducting numerical computation  with {\sc{Bocop}} for a number of links $N>5$ is currently beyond its limitations of memory allocation. There is however no such limitations for the solution of the small amplitude version of the optimal control problem \eqref{eq:phi_star_final} presented here.

\section{Conclusion}
In this work, we have studied optimal periodic strokes of multi-link micro-swimmers that achieve a given prescribed displacement in a given time with minimum energy. Exploiting the linearity of Stokes flows and geometric symmetries of the swimmer, the optimization has been formulated as a constrained variational problem, where leading-order expansion of the dynamics leads to the optimal solution of an eigenvalue problem. Remarkably, it is proven that energy-optimal strokes for $N$-link swimmers reduce to ellipses lying in a two-dimensional subspace of the space of shapes. 

For large $N$, the optimal strokes become traveling waves with the shortest possible wavelength of three links, in agreement with the observation made in Taylor's classic work \cite{Taylor51}. A noticeable difference from \cite{Taylor51} due to the finite length of the swimmer is the slightly non-uniform distribution of energy-optimal wave amplitudes, which decay symmetrically from swimmer's center towards its ends. 

Numerical optimization for the fully nonlinear problem, in which the amplitude of the prescribed displacements and the excursions of the joint angles are allowed to be large, 
is obtained using the optimization toolbox  {\sc{Bocop}} for multi-link swimmers with three and five links.
When the prescribed displacements are small, the numerical results show excellent agreement with those obtained using the small amplitude approximation.
When the imposed displacements are large, the picture changes significantly.
For Purcell's 3-link swimmer, we obtain non-convex closed curves similar to the ones previously  obtained by Tam and Hosoi for maximizing displacement or Lighthill's efficiency. 
For the 5-link swimmer, we obtain non-planar loops of complex and intriguing geometry. 
A more precise characterization of the properties of  these highly-dimensional closed loops seems an interesting open problem.
For both $N=3$ and $N=5$, all these complex shapes converge to the planar ellipses of the small amplitude approximation when the prescribed displacements become small.

Possible directions for future extension of this research are comparison with measured strokes of biological swimming microorganisms as in \cite{berman2013undulatory,Rossi2017Kinematics}, generalization of the model to include elasticity as in \cite{Montino,spagnolie2010optimal,passov2012dynamics,cicconofri2016motion}, and the study of optimal control of magnetically-actuated microswimmers \cite{alouges2015soft}.
 
\section*{Acknowledgements}

OW wishes to thank the Technion-France Association for their support in his research internship visit to Paris in 2015, in which this work has been initiated.
We gratefully acknowledge support by the Israel Science Foundation (OW and YO, grant no. 567/14) and by the European Research Council (ADS, grant no. 340685-MicroMotility).
We all wish to warmly thank Dr. Pierre Martinon for his help and technical support with the {\sc{Bocop}} software.


\section*{Appendix - scaling laws for large N}
\begin{figure}
\centering
\begin{subfigure}{0.45\textwidth}
\includegraphics[width=\textwidth]{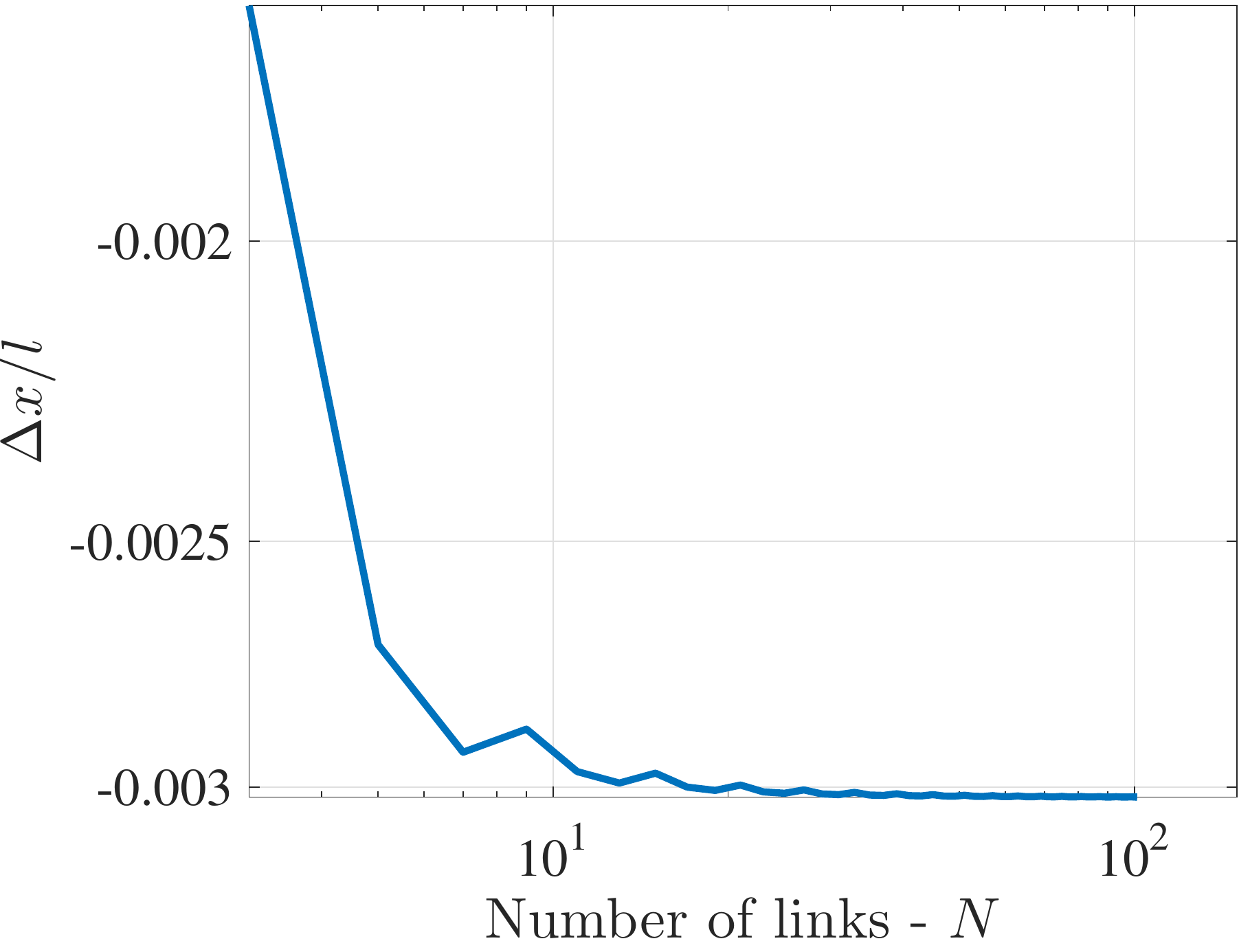}
\caption{}
\label{fig:apndx_x_n_const_l}
\end{subfigure}
\begin{subfigure}{0.45\textwidth}
\includegraphics[width=\textwidth]{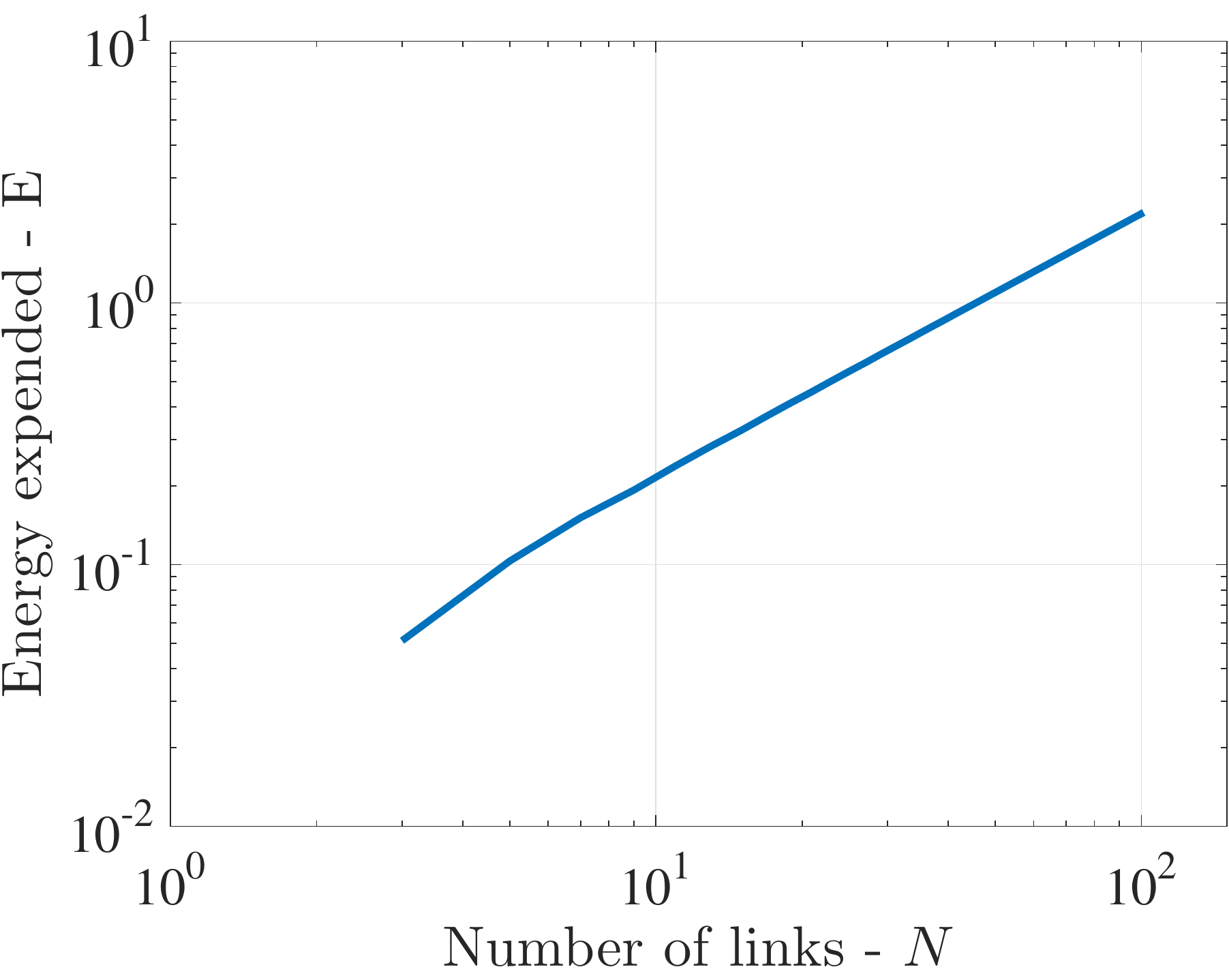}
\caption{}
\label{fig:apndx_E_n_const_l}
\end{subfigure}
 \caption{Log-log plots of (a) displacement $\Delta x$ and (b) total energy E as a function of links number $N$, under the traveling-wave input \eqref{eq.wave} for a chain of $N$ links with equal lengths $l=1$.}
 \label{Fig:function_of_N_const_l}
 \end{figure}

In order to study scaling of the displacement and energy for large $N$, we now consider a swimmer with $N$ links of fixed lengths $l=1$, under the gait of travelling wave:
\begin{equation} \label{eq.wave}
\phi_k(t)=\vep \sin(\omega t + k \Delta \varphi) \mbox{, for } k=1 \ldots N \mbox{, where } \omega = 2 \pi / T.
\end{equation}
We consider gaits with fixed amplitude $\vep$ and phase difference of $\Delta \varphi=2 \pi /3$. 
Using our formulation based on resistive-force theory \eqref{eq.rft}, we numerically calculate the displacement $\Delta x$ and energy expenditure $E$ along a cycle, as a function of $N$. The results are shown on log-log scale plots in Figures \ref{fig:apndx_x_n_const_l} and \ref{fig:apndx_E_n_const_l}. It can be seen that for large $N$, the displacement $\Delta x$ converges to a constant. This is analogous to the ``infinite sheet'' limit of Taylor's net swimming speed \cite{Taylor51}. In addition, $\Delta x$ is linear in $l$ and scales quadratically with the stroke amplitude $\vep$ as in \cite{Taylor51}. The energy expenditure $E$, however, grows linearly with $N$. This is because for large $N$, the interaction between links becomes negligible and each link contributes an equal amount of viscous dissipation. In addition, computation for fixed $N$ and varying links' length $l$ reveals that the energy expenditure scales cubically, as $l^3$ (plot not shown). This is explained by the following observation. The mechanical power dissipation of each link scales as $P_i=f_i v_i$ where  $f_i$ is the viscous drag force and $v_i$ is the link's linear velocity. The viscous force $f_i$ scales as $f_i=R v_i$ where $R$ is a viscous resistance coefficient (cf. \cite{BrennerHappel65,cox1970,GrayHancock55}). The link's velocity scales as $v_i \sim \vep l /T$, while the resistance coefficient $R$ scales linearly with the link's length $l$. Since total energy dissipation scales as $N P_i T$, we deduce the following scaling laws for large $N$:
\begin{equation} \label{eq.scaling}
\Delta x \sim \vep^2 l, \;\;\;  E \sim \vep^2 N l^3 /T .
\end{equation}
Consider now the case where the required displacement $\Delta x$ as well as the \textit{total length} of the swimmer $L=Nl$, are held fixed while $N$ is varied. That is, the links' length scales as $l=L/N$. The scaling laws in \eqref{eq.scaling} will now change to
\begin{equation} \label{eq.scalingL}
\Delta x \sim \vep^2 \frac{L}{N}, \;\;\;  E \sim \vep^2 \frac{L^3}{N^2 T}  .
\end{equation}
Fixing the displacement $\Delta x$, it is deduced from \eqref{eq.scaling} that for large $N$, the amplitude $\vep$ 
increases as $\sqrt{\frac{N\Delta x}{L}}$.
Moreover, the energy scales as $E \sim \frac{\Delta x L^2}{NT}$. This scaling relation explains the decay rate of $E^*$ in Figure \ref{Fig:constant_L}.

\end{document}